\documentclass[preprint,10pt]{aastex}
\usepackage{hyperref}
\usepackage{natbib}

\usepackage{amsmath}

\slugcomment{submitted to AJ 20170512; revised 20171018}
\slugcomment{accepted to AJ 20171204}

\title{Robo-AO Discovery and Basic Characterization of Wide Multiple Star Systems in the Pleiades, Praesepe, and NGC 2264 Clusters}

\author{Lynne A.\ Hillenbrand\altaffilmark{1}, 
Celia \ Zhang\altaffilmark{1}, 
Reed L. \ Riddle\altaffilmark{2}, 
Christoph Baranec\altaffilmark{3}, 
Carl Ziegler\altaffilmark{4}, Nicholas M. Law\altaffilmark{4},
John Stauffer\altaffilmark{5}
}
\altaffiltext{1}{Department of Astronomy, California Institute of Technology, Pasadena, CA 91125, USA; {\tt lah@astro.caltech.edu}}
\altaffiltext{2}{Caltech Optical Observatories, California Institute of Technology, Pasadena, CA 91125, USA} 
\altaffiltext{3}{ Institute for Astronomy, University of Hawai\'i at Ma\~noa,
Hilo, Hawai\'i 96720-2700, USA}
\altaffiltext{4}{Division of Physics and Astronomy, University of North Carolina at Chapel Hill, Chapel Hill, NC 27599-3255, USA}
\altaffiltext{5}{Spitzer Science Center, California Institute of Technology, Pasadena, CA 91125, USA}

\begin{document}

\begin{abstract}
We identify and roughly characterize 66 candidate binary star systems 
in the Pleiades, Praesepe, and NGC 2264 star clusters 
based on robotic adaptive optics imaging data obtained using Robo-AO
at the Palomar 60" telescope.
Only $\sim$10\% of our imaged pairs were previously known.  
We detect companions at red optical wavelengths having physical separations 
ranging from a few tens to a few thousand AU.  A 3-sigma contrast curve 
generated for each final image provides upper limits 
to the brightness ratios for any undetected putative companions.  
The observations are sensitive to companions with maximum contrast $\sim$6$^m$
at larger separations.  At smaller separations, the mean (best) 
raw contrast at 2\arcsec\ is 3.8$^m$ (6$^m$), 
at 1\arcsec\ is 3.0$^m$ (4.5$^m$), and at 0.5\arcsec\ is 1.9$^m$ (3$^m$). 
PSF subtraction can recover close to the full contrast in to the closer separations.
For detected candidate binary pairs,  
we report separations, position angles, and relative magnitudes. 
Theoretical isochrones appropriate to the Pleiades and Praesepe clusters 
are then used to determine the corresponding binary mass ratios, 
which range from 0.2-0.9 in $q=m_2/m_1$. 
For our sample of roughly solar-mass (FGK type) stars in NGC 2264 and
sub-solar-mass (K and early M-type) primaries in the Pleiades and Praesepe,
the overall binary frequency is measured at $\sim$15.5\% $\pm$ 2\%. 
However, this value should be considered a lower limit to the true binary fraction 
within the specified separation and mass ratio ranges in these clusters,
given that complex and uncertain corrections for sensitivity and completeness 
have not been applied.
\end{abstract}


\section{Introduction}

More than half of all stars are found in multiple star systems. 
As reviewed by e.g. Goodwin et al. (2007), Duchene \& Kraus (2013), 
and Reipurth et al. (2014), stellar multiplicity properties 
appear to be set within the first few million years of a star's life. 
Binary frequency is observed to increase with primary star mass, 
from $<$25\% for stars near the hydrogen-burning limit to $>$90\% for stars
near the top of the initial mass function.  The similar distributions
in mass ratio and semi-major axis for both main sequence and pre-main sequence 
binaries, suggests that the same formation processes occur over 
varying core masses, and that mild core fragmentation of collapsing
gas clouds is the most appropriate theory for multiple-system formation
(as reviewed by Boss, 1995; see also Offner et al. 2010). 

In addition to having approximately the same distance from Earth, 
members of a star cluster form at the same time, from the same gas cloud, 
and have similar age and chemical composition. Thus, studying members of
young clusters provides insight into star formation and evolution. 
A ``top-down" theory describing the formation of stars in clusters 
invokes cloud compression and fragmentation through shocks due to
supersonic turbulence.  Approximately Jeans-mass fragments then collapse 
into single or, under further fragmentation, into binary and higher order 
multiple star systems that may then undergo further subsequent 
dynamical evolution, including binary disruption and/or capture, 
as reviewed by e.g. Goodwin \& Kroupa (2005) and Bodenheimer (2011). 
The ``bottom-up" theory (e.g. Shu et al. 1987) describes 
turbulence shocked gas that collapses directly into cores 
which become opaque to the radiation generated from conversion of 
gravitational energy, and continue to accrete gas. These objects evolve
from near the $\sim 10^{-3}$ M$_\odot$ opacity-limited fragmentation limit, 
all the way up to become stars.  In this model, binaries at $\lesssim$100 AU 
can form at later times through disk fragmentation, if the disk is cool enough.

Binary star characteristics such as multiplicity frequency, 
separation distribution, and mass ratio distribution can support or refute 
different elements of the various star and multiple star formation theories,
with trends in these distributions as a function of primary star mass
particularly important to quantify. 
Here, we examine wide-separation multiplicity in the 
Pleiades, Praesepe, and NGC 2264 clusters.
We discuss the clusters in order of increasing distance because 
our observations are limited by both angular resolution (limiting detection 
of companions at constant photometric sensitivity), and by sensitivity 
(limiting the measurement of flux ratios at constant spatial resolution). 
Both resolution and sensitivity improve for closer targets.

The Pleiades cluster is one of the youngest and closest populous star clusters. 
It is well-studied and has had its constituent stars cataloged extensively
(Rebull et al. 2016). 
The cluster has $\sim$1500 known members that are readily identified 
due to a significant common proper motion compared to background stars. 
From lithium depletion boundary methods, 
the cluster age was determined to be 125 $\pm$ 8 Myr (Stauffer et al. 1998b) 
while the mean distance is 136 $\pm$ 1 pc (Melis et al. 2014) and the mean
extinction is often quoted as $A_V=0.15$ mag. 
Praesepe also has been extensively cataloged (Rebull et al. 2017), 
aided by its distinct proper motion (e.g. Adams et al., 2002; Wang et al. 2014).
The $\sim1000$ members of Praesepe are intermediate in age at 757$\pm$36 Myr 
and have a mean distance of 179$\pm$2 pc (Gaspar et al. 2009) and negligible reddening. 
NGC 2264 is a young $\sim$3 Myr cluster that is significantly further away,
having a mean distance somewhere between $\sim$740 pc (Kamezaki et al 2014) 
and $\sim$913 pc (Baxter et al. 2009) 
{with a range of extinction among the} around 1500 known members. 
The cluster is a popular target for young star studies because it has 
moderately low extinction along the line of sight, and is next to a molecular
cloud complex that reduces contamination from background stars.

As for exoplanet populations work,
stellar multiplicity studies are conducted using techniques that sample
different portions of the mass ratio $q=m_2/m_1 $ versus semi-major axis, $a$,
parameter space.
Spectroscopic monitoring can detect radial velocity variations from either just one or both stars.
Photometric monitoring searches for eclipses. 
Both methods sample close-in orbits, or small values of $a$, more readily 
than larger values of $a$.  Direct imaging, the method employed here,
samples only larger $a$ values.  All detection methods are most sensitive 
to binaries having small differences in mass / size / brightness, 
and become less sensitive towards lower mass, smaller, and fainter companions.
Previous multiplicity work on the particular clusters we have investigated 
includes: radial velocity, eclipse, and direct imaging studies,
as well as photometric identification of binaries.

Among both Pleiades and Praesepe cluster members, the binary fraction is 
found to be higher for the more concentrated members
(Raboud \& Mermilliod; 1998, 1999), which is attributed
to the general trend of increased multiplicity towards higher mass stars
that are more centrally concentrated than lower mass stars, 
perhaps due to the effect that their multiplicity increases 
the system mass and hence shortens the time scale for mass segregation (van Leeuwen, 1983).

In the Pleiades, Bettis (1975), Jaschek (1976),
Stauffer et al. (1984), Pinfield et al.  (2003), and Lodieu et al. (2012) 
assessed multiplicity based on photometric binary 
candidates, collectively covering A through L spectral types. Pinfield et al.
(2003) concluded that the binary fraction increases towards lower masses
while Lodieu et al. (2012) found a brown dwarf binary frequency using this
technique of about 24\% within 100 AU.  These results have significant tension
(at the factor of 2-3 level) with the direct imaging results covering the
same mass and separation ranges that are discussed below.
Previous and new radial velocity measurements of BAFG stars were used by 
Mermilliod et al. (1992, 1997) and Raboud \& Mermilliod (1998) to characterize
multiplity, resulting in a $\sim$25\% spectroscopic binary fraction,
consistent with the field star population.  
Adaptive optics direct imaging of G and K dwarfs 
was used by Bouvier et al. (1997) to estimate a binary fraction of 28$\pm$4\%
between 11 and 910 AU, consistent with the field star population. 
Martin et al. (2000), Bouy et al. (2006), and Garcia et al. (2015)
all searched using HST for binaries among small samples of brown dwarfs, 
finding results consistent with the low binary fraction 
of $\sim$15\% or less in the field for the mass and separation range.
Notably, the majority of their newly resolved systems had been identified
previously as photometric binary candidates (Raboud \& Mermilliod; 1998). 
Richichi et al. (2012) identified several Pleiades binaries 
from lunar occulation observations.

In Praesepe, Bettis (1975), Jaschek (1976), and Pinfield et al. (2003)
identified photometric binary candidates in this cluster as well. 
Bolte (1991) confirmed the early candidates as true binaries via spectroscopy.
Later, Boudreault et al. (2012) and Khalaj \& Baumgardt (2014) 
addressed multiplicity in a statistical sense with the latter authors 
measuring a multiple fraction of 8.5$\pm$1.6\% and then claiming
a true binary$+$triple fraction of 35\%.  
Radial velocity measurements were used by Mermilliod \& Mayor (1999)
to measure spectroscopic binary orbits for FGK stars.
A direct imaging search for multiplicity was conducted by 
Bouvier et al  (2001) who estimate among G and K stars a binary fraction 
of 25$\pm$5\% between 15 and 600 AU, in agreement with the field. 
Patience et al. (2002) surveyed B through M stars and 
found a smaller binary fraction over the same separation range, 
but also had a smaller sample. 
Peterson et al. (1984, 1989) identified several new Praesepe binaries 
from lunar occulation observations.

In NGC 2264, there have been few dedicated binary studies. 
Recently, Gillen (2015, 2017) has conducted a successful search 
for eclipsing systems and Kounkel et al. (2016) has identified 
spectroscopic binaries.

The overall conclusion from the above, and other papers on binarity in clusters,
is that the dense cluster binary statistics are similar to the field
star population binary statistics. This is in contrast to the results 
for young loose associations, which appear to have higher binary fractions,
and lends support to the idea that the majority of the field star population
formed in clusters rather than looser associations.  An alternate hypothesis is 
that some binaries in young associations breakup during their pre-main sequence evolution.

\section{Sample Selection for Binary Search}
\label{sec:sample}

The input samples for our adaptive optics direct imaging companion search
were selected as described below.  A primary consideration was the availability
of high cadence and high precision photometric datasets 
(either available at the time, or pending) for likely members of each cluster.
All potential targets were within the brightness range J$\approx$10-13.5$^m$, 
selected as such via 2MASS photometry.  
From the input samples, the robotic scheduler for Robo-AO (Riddle et al. 2014) 
chose the actual targets of observation, as described below.

In NGC 2264, we included the 100 brightest classical T Tauri stars and
the 100 brightest weak-line T Tauri stars having time series data from CoRoT.
The CoRoT sample selection and results are described in e.g. 
Affer et al. (2013), Cody et al. (2014), Sousa et al. (2016), 
Lanza et al. (2016), Venuti et al. (2017), and Guarcello et al. (2017).

For the Pleiades and Praesepe clusters, the stars come from the $K2$ open clusters
investigation of J. Stauffer (program IDs GO4032 and GO5032) 
which aimed to completely survey the bona fide members of these two clusters with high-precision, high-cadence optical photometry.
The $K2$ time series sample selection and resulting data are described in 
detail in Rebull et al. (2016) for the Pleiades, and Rebull et al. (2017) for Praesepe.
Pleiades stars were selected from the samples provided in Stauffer et al. (2007)
and Bouy et al. (2015). Praesepe stars were compiled from the work of 
Jones \& Cudworth (1983), Jones \& Stauffer (1991),
Klein-Wassink (1927), and Kraus \& Hillenbrand (2007).

We obtained Robo-AO data for a total of 120 NGC 2264 members, 
212 Pleiades members, and 108 Praesepe members. 
Knowledge regarding stellar multiplicity or lack thereof 
can inform the CoRoT and K2 lightcurve analysis.

\section{Robo-AO}
\label{sec:data}

\subsection{Hardware and Operation}

Robo-AO (Baranec et al. 2014) was mounted on the Palomar 60-inch (1.5m) 
telescope before being moved in 2015 to the 
Kitt Peak 2.1m telescope (Salama et al. 2016; Jensen-Clem et al. 2017). 
At Palomar, the instrument was capable of imaging more than 200 objects per night.

Robo-AO is the first autonomous laser AO system (Baranec et al. 2014).  It uses a 10-W
ultraviolet laser for a guide star, which releases a 35-nanosecond laser pulse every 100
microseconds, and records the Rayleigh-scattered, returning photons to determine the
correction (Riddle et al. 2015).  The laser pulse is focused at 10 km.
Wavefront aberrations are sampled at 1.2kHz which is sufficient to measure and correct the
wavefront errors of the Palomar 1.5m telescope.  
The 44\arcsec\ field-of-view of the Robo-AO visible camera is contained within
the well-corrected diffraction-limited image area. 

Besides the laser-launch system, the instrument consists of:
a set of support electronics; a Cassegrain instrument package 
that houses a high speed electro-optical shutter, wavefront sensor, wavefront
corrector, science instrument and calibration sources; and a single computer that controls
the entire system. A master sequencer to control hardware subsystems 
creates an efficient overall observing system. 

A queue scheduling program selects Robo-AO targets, 
optimizing the desired science targets among the practical constraints. 
A robot sequencer points the
telescope and configures the associated system. A laser acquisition process 
involves a search algorithm to move the uplink steering mirror; 
the entire instrument configuration takes less than 30 seconds after telescope
slew. During observations, telemetry is used to maintain focus and detect 
drops in laser return to continually adapt to conditions. Data 
are stored and processed in a computer system separate from the instrument 
control (Baranec et al. 2014).  
An Andor iXon DU-888 EMCCD images the science field at 8.6Hz 
(116 msec per individual frame) with the images saved in data cubes for later processing.

The Robo-AO system can be used to build 
a sizable sample of moderate-contrast diffraction-limited images of stars 
within just a few nights.

\subsection{Data Acquisition and Image Processing}

Images of 446 targets in our three clusters were collected 
using the Robo-AO system on 2014, November 7-11, and 2015, March 3. 
Either an SDSS i' filter or an LP600 long pass filter with a 600 nm cut-on 
was used, the latter for sources fainter than R $\approx$ 13$^m$ 
given the increased filter breadth but similar effective wavelength.

The observing sequence for a single target, described in detail in Baranec et al. (2014), begins with a queue scheduling program that optimizes among scientific priority, slew time, telescope limits, prior observing attempts, and laser-satellite avoidance windows. The science camera, laser, and adaptive optics system are configured as the telescope slews.  Once pointed at the new target, the laser is acquired with a search algorithm moving a steering mirror, the adaptive optics system is started, and an observation is performed with no adaptive optics correction to estimate seeing conditions. Once the laser is acquired, the adaptive optics correction is started, removing residual atmospheric wavefront aberrations at 100Hz using a 12 x 12 actuator deformable mirror. 

Total exposure times were either 120 sec or 300 sec, depending on source brightness.
The raw data files consist of multiple data cubes of visible camera frames generated at 8.6Hz. When a cube reaches a size of 1GB (about 256 frames) it is closed and a new cube generated.  In a 120 (300) sec exposure, about 1032 (2580) individual frames are generated.  

The Robo-AO image processing pipeline is described in detail in Law et al. (2014ab).
Each of our images was dark-subtracted, flat-fielded, and tip-tilt-corrected. 
Tip-tilt image motion was corrected using an object brighter than $\sim$16$^m$
in the science image field, with a post-facto shift-and-add routine in the pipeline; in the case of our observations, the tip-tilt star was the main target.  
The images from each data cube were then stacked into a composite image with
an up-sampled plate scale of 21.55 mas/pixel\footnote{This is half the pixel scale 
quoted in Riddle et al. (2015) since the images in the present data set
are oversampled by a factor of two.}. 
The data pipeline can select only a percentage of the best quality frames for inclusion in the final image (as is done with so-called ``lucky imaging"), but in practice all Robo-AO frames are used in producing the final output science image.

Image cutouts of $400 \times 400$ pixel$^2$
(about 8.6\arcsec\ by 8.6\arcsec), centered on each target star, were made for the subsequent analysis steps.  Then 
a locally optimized PSF was created for each target from at least 20 sources 
observed nearby in time and airmass to the target.  
The reference PSF images were other science images taken temporally close to the science observation (within 1-2 hours); any drift in the PSF during a night should be slow.  
The target PSF is modelled using a linear combination of the reference PSFs (employing the LOCI algorithm; Lafreniere et al. 2007).  If a reference PSF does not correlate with the target PSF, the algorithm does not include it in the model.

\begin{figure}[htpb]
\vskip-10truein
\includegraphics[width=2.0\textwidth]{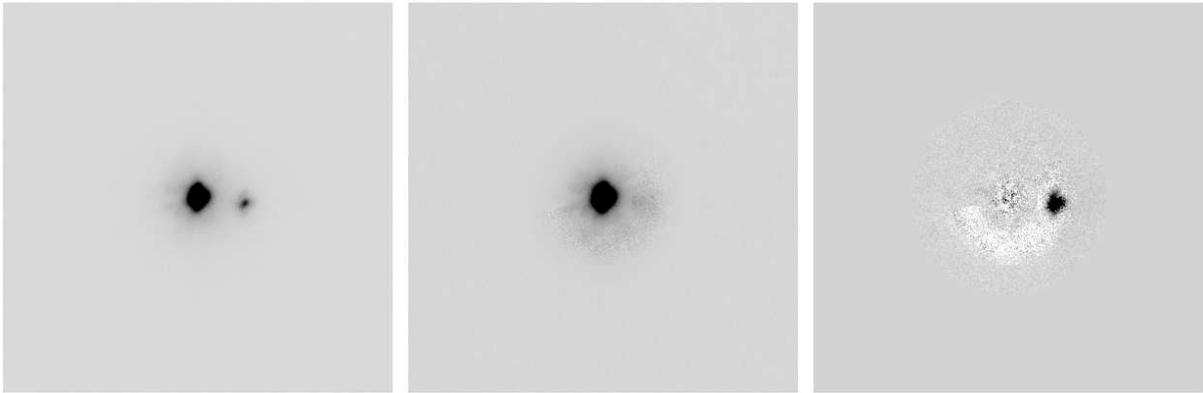}
\caption{
For AK IV-314, a binary system, shown are the stacked image cutout (left),
the PSF image assembled from images surrounding the leftmost image in time (center), 
and the PSF-subtracted remainder image (right) 
that isolates the secondary star. 
}
\label{fig:images}
\end{figure}

This empirical PSF was subtracted from the target stacked image cutout, and 
both it and the remainder image were saved along with the 
cutout. Figure~\ref{fig:images} illustrates a PSF-subtraction sequence.

We note that in some images, artifacts from the data acquisition and
processing produced apparent ``triple systems" that are identifiable because
of their linear alignment on the image.  
Software was used to remove this effect, 
but it failed for some targets ($<1$\% of the sample) and they had to be
discarded from further analysis.

Each good final image stack contains either only a single source, 
or a detected candidate binary pair. 
For most of the targets, the stacked image cutout allowed us to identify, 
measure separations and position angles, and photometer the binary components. 
However, in cases where the secondary star is at high enough contrast or 
intrinsically too faint to be detected in the initial cutout, 
the PSF-subtracted image can be used to determine multiplicity and 
measure the properties of the secondary.

\subsection{Image Quality Analysis}
\label{sec:psf}

Image quality was used to assess the performance of the AO system and to determine the significance of the point sources detected in an observation.

\begin{figure}[htpb]
\centering
\includegraphics[width=0.49\textwidth]{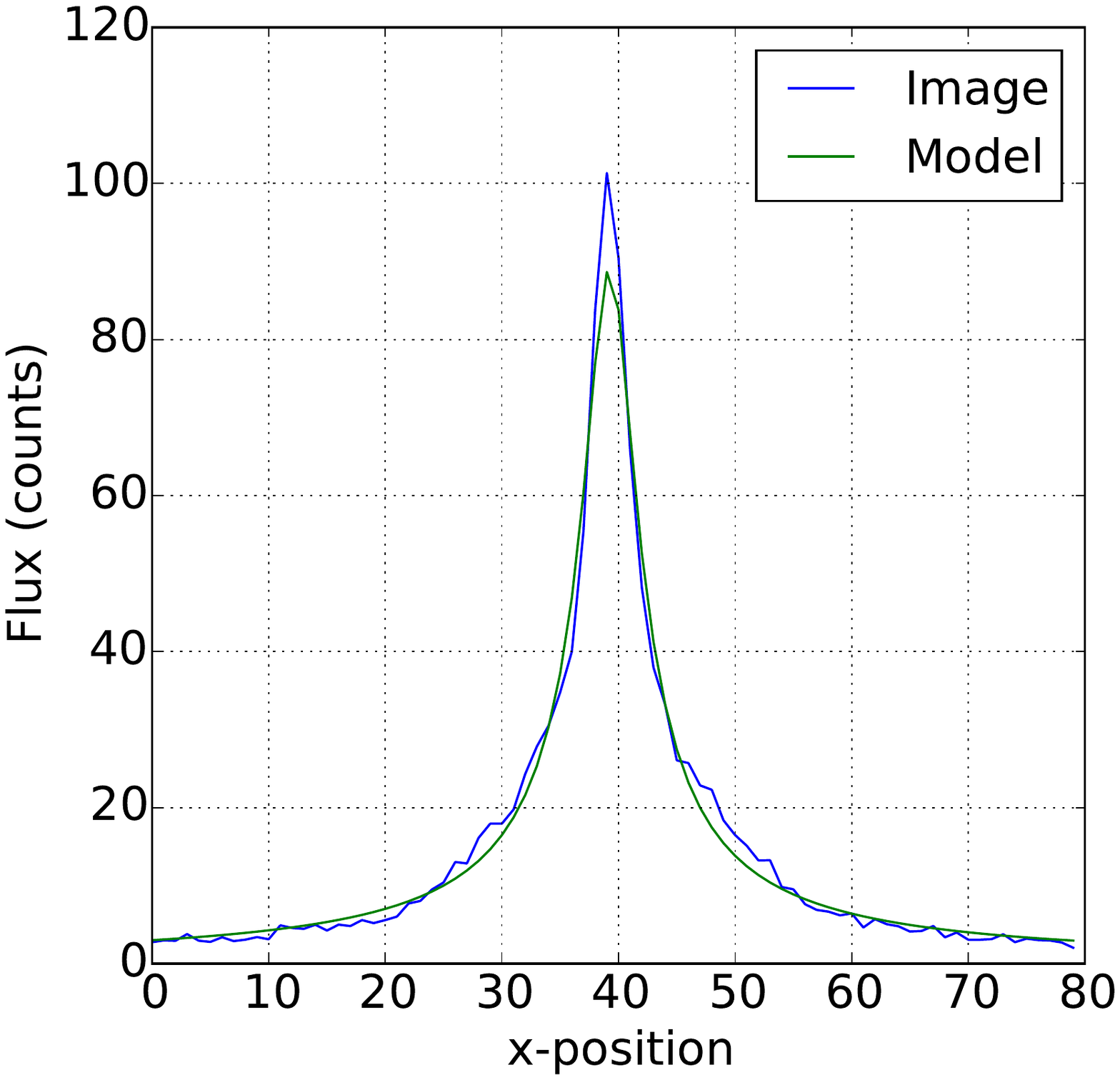}
\includegraphics[width=0.49\textwidth]{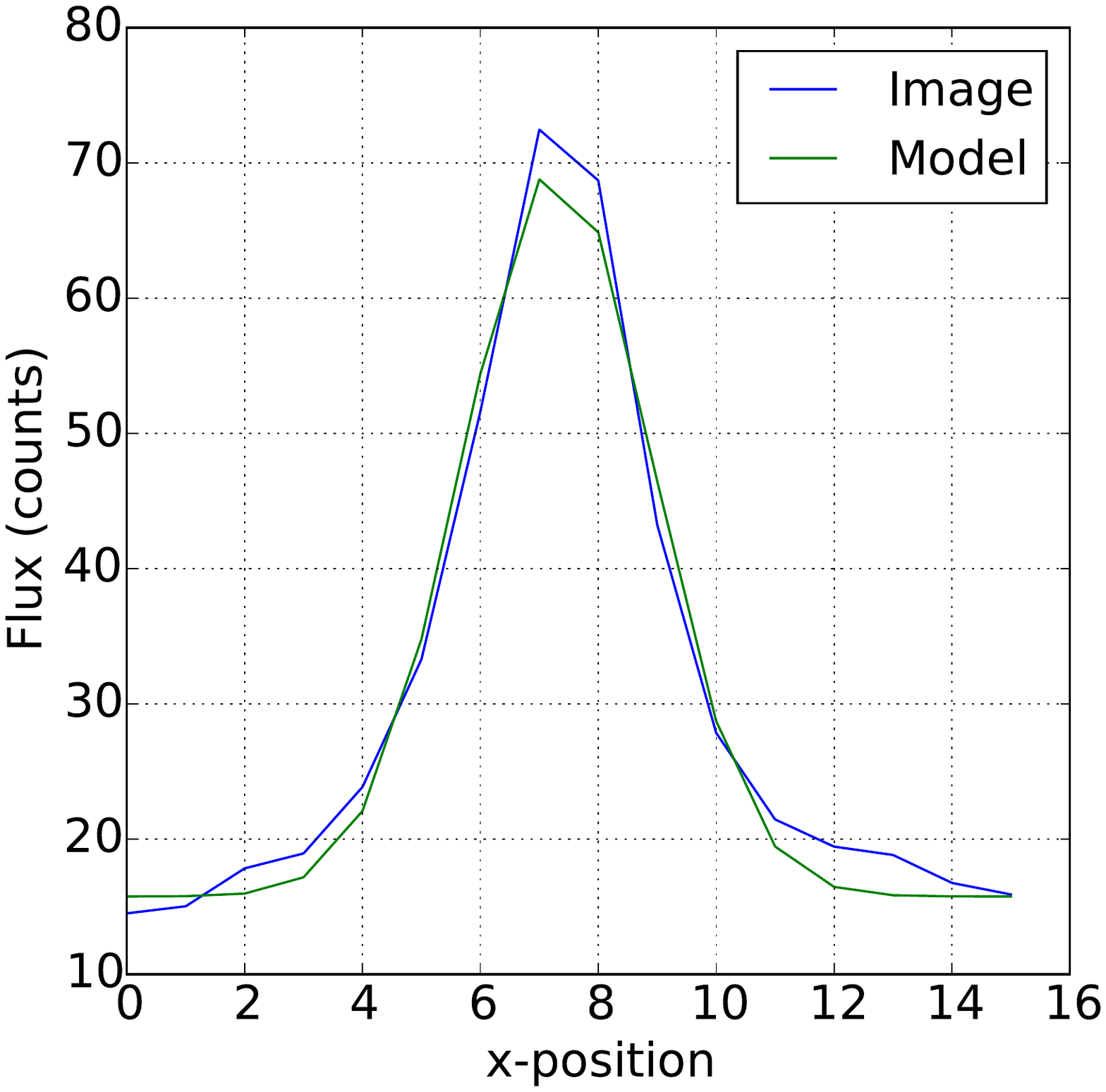}
\vskip-0.5truein
\caption{ 
Example PSF models, evaluated at whole pixel values, and matched to cuts 
across Robo-AO image data. 
Left: 2D Moffat model of source BPL 167.  
Since the peak of the model is not located at the center of a pixel, 
and partial pixel values are interpolated in the plot,
the peak is not well-represented graphically but is better estimated numerically.
This model matches the image wings, but slightly overestimates the FWHM 
in the image core. 
Right: 2D Gaussian model of source CSIMon-0722. 
Relative to the Moffat model, the Gaussian model better matches the FWHM 
in the image core, but is worse in the image wings. 
}
\label{fig:psfmodels}
\end{figure}

We modeled the primary point-spread function (PSF) on each image
to assess image quality and to monitor its variation 
across the data set.  High-quality images have smaller FWHMs 
and are produced when the AO correction delivers a diffraction-limited PSF. 
Larger FWHM images occur when the PSF is dominated by uncorrected 
atmospheric turbulence, or poor seeing.  Due to the image registration
process, in cases of exceptionally poor correction, a single bright 
central pixel results from the stacking of seeing limited images 
(see discussion in Law et al. 2014).  To find the FWHM of each stacked
cutout image, we fit curves with different functional forms to the flux
profile; examples are shown in Figure~\ref{fig:psfmodels}.

We first fit two-dimensional Moffat functions, of the form 
$$ f(x,y)=A(1+ {{(x-x_o)^2 + (y-y_o)^2}\over{\gamma^2}})^{-\alpha}$$
where $x_o$ and $y_o$ describe the centroid position, 
$x$ and $y$ are the spatial coordinates on the image, 
and $A$, $\gamma$, and $\alpha$ are free parameters.  
Following the profile fit, its FWHM was defined as
$$ FWHM=2\gamma (2^{1/\alpha}-1)^{0.5}.$$
In any given image, the number of pixels at essentially the background 
sky level far exceeds the number of pixels with significant amounts of flux, 
so we restricted the fitting box size around the primary.  However, the Moffat 
fit failed to produce sensible results if the wings of the PSF 
were not entirely captured. We thus began with a $10\times 10$ pixel 
($0.22\arcsec \times 0.22\arcsec$) box and incremented the box side length 
in 5 pixel steps, until a minimum FWHM was found. 
Based on visual examination, the Moffat fits tended to overestimate the FWHM. 
For some very faint sources, however, the FWHM was underestimated since the 
error term used to constrain the Moffat fit did not interpolate between pixels, 
greatly exaggerating the image peak. 

We next fit two-dimensional Gaussian functions, of the form
$$g(x,y)= Ae^{- {{(x-x_o )^2+(y-y_o )^2}\over{2\sigma^2 }}}+C$$
with free parameters $A, C, \sigma$.  The FWHM is then
$$FWHM = 2(2 log(2))^{0.5}\sigma.$$
As for the Moffat fitting, the box size for the Gaussian fit was restricted.
An additional challenge for the Gaussian fitting was that too few points 
could be considered when fitting the model. We thus began with a fitting box of length 
four times the Moffat-derived FWHM, and decremented it 1 pixel at a time 
until the RMS of the residual error of the centroid coordinate and its 
eight neighbors was below 0.1; we limited the box size to a minimum 
of $10\times 10$ pixels.

\begin{figure}[htpb]
\centering
\includegraphics[width=0.49\textwidth]{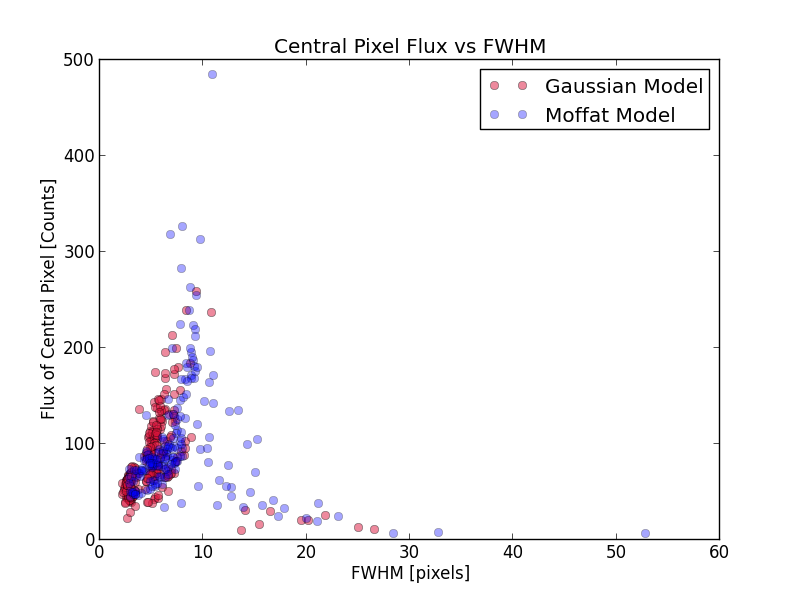}
\includegraphics[width=0.49\textwidth]{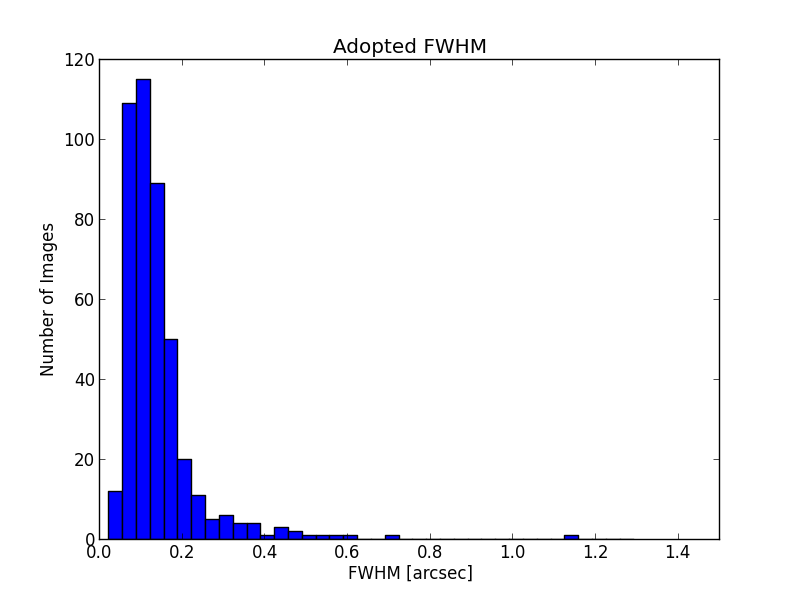}
\caption{
FWHM relation to source brightness, for both the Gaussian and the Moffat functional fits 
(left panel), and final adopted FWHM value as given in Table~\ref{tab:sources} (right panel).
}
\label{fig:fwhm}
\end{figure}



Figure~\ref{fig:fwhm} illustrates the relationship between FWHM and source brightness 
derived under each model; Moffat fits are systematically larger than Gaussian fits. 
Figure~\ref{fig:fwhm} also illustrates the FWHM distribution in arcsec and demonstrates
that the images are, for the most part, diffraction limited.
The vast majority of estimated FWHMs are 3-10 pixels, or 0.07-0.22\arcsec\ with a 
minority in a tail extending to outliers as high as 50 pixels, or 1.1\arcsec, 
which implies negligible AO corection for these several objects.  However, 
given that the diffraction limit of the Palomar 60-inch telescope at Robo-AO wavelengths 
is $\lambda/D = 0.25 \times 0.75\mu m /1.52 m = 0.12$\arcsec, or $\sim$5.5 pixels,
FWHM values less than this (formally $1.028 \times \lambda/D$) are spurious;
these are attributed to the image stacking process, which for low signal-to-noise 
sources can enhance a central peak noise spike.  
The underestimated FWHM values are not important for the photometry, 
which used a diffraction-limited aperture.
The rise in FWHM with source brightness is expected, 
as the size of the PSF would increase.  However, there is also a group of 
dimmer stars with unexpectedly large FWHMs, which we attribute to 
the AO correction not working as well due to the faintness of the stars, 
poor seeing, or possibly telescope motions -- 
all of which can spread out the PSF of fainter objects. 

The shape of the PSF varies as observing conditions and equipment performance 
change. It is for this reason that we combine local-in-time images in creating
the empirical PSFs for subtraction from individual sources  
(Figure~\ref{fig:images}). Consequently, the theoretical model that most 
effectively approximates the PSF also changes over time.  
For each final image set, we therefore chose whichever model 
(Moffat or Gaussian) produced a smaller central-9-pixel RMS residual error term,
in order to define the FWHM.   
The mean and median FWHM values are 6.45 and 5.51 pixels or 0.14 and 0.12\arcsec. 
For the high FWHM, faint sources, 
the Moffat fit yields better results.  Similarly, for the low FWHM, 
bright sources, the Moffat fit also yields better results.

The FWHM values are reported in Table~\ref{tab:sources} and, as expected,
are anti-correlated with the contrast sensitivity limits that are presented below.

\section{Analysis of Detected Binaries}

To determine the existence of binarity, 
we first used the DS9 V7.3.2 software\footnote{http://ds9.si.edu/} 
to examine the final image of each target.  Overall brightness and contrast was varied, and contour plots were produced in identifying the candidate binaries. 

\subsection{Astrometry}
DS9 was also used to roughly gauge the 
relative locations of the constituent stars in a pair. 
For improved precision, we used the Aperture Photometry Tool (APT; Laher, 2015) 
software\footnote{http://www.aperturephotometry.org/} to determine the
pixel centroids of each star in the initial cutout images.
APT operates via a GUI with which users may manually place an aperture on a source. 
APT then uses iterative methods to compute centroid positions and, as discussed in the next section, aperture fluxes (Laher et al. 2012).

Before we could calculate the binary separations and position angles, 
we had to correct each primary centroid for image distortion (Riddle et al., 2015). 
Each cutout image has a reference coordinate with respect to the earlier stage full image
taken by the Robo-AO instrument, enabling us to apply the published distortion correction.
We then calculated the separation and rotation for each binary 
using the corrected centroids and knowledge of array orientation relative to true north.
Finally, pixel separations were converted to physical separations in AU adopting 
the plate scale sampling of 21.55 mas and respective distances of 136, 175, and 740 pc 
for the Pleiades, Praesepe, and NGC 2264 clusters.

An appropriate typical error for the measured separations is $\sim$0.02\arcsec,
and for the measured position angles is $\sim$0.1-0.3 deg; the true error in the latter
is dominated by an additional uncertainty in the instrument orientation, perhaps up to
1.5 degree based on repeated calibrations using globular cluster fields (Baranec et al, 2016).

\subsection{Photometry}

By convention, the star that is visibly brighter is designated as the
primary, and its companion is the secondary.  To obtain the difference 
in brightness between the two stars of each binary system, we
used APT to measure the magnitude of each source inside an aperture,
along with an uncertainty.
Aperture corrections are not needed, since we are measuring
relative magnitudes and the PSFs of each star are the same, 
given that they appear in the same image.

We used Model F for background correction within APT, which
is a non-annulus based local estimate for the sky background
considering a grid size of 64 pixels and window size of 129 pixels. The other five sky
subtraction models APT offers take the mean, median, or mode of a manually
placed sky-annulus, set the sky value manually, or apply no sky subtraction. We
decided against using any of the models that are based on a sky annulus since the target stars are very large
with respect to the image size, and of varying size and FWHM, 
so determining sky annulus radii introduced an unnecessarily arbitrary element to the analysis.
For very faint secondary stars or binary systems with small separations, the algorithm
used by APT to obtain centroids could not pinpoint the secondaries on the initial 
cutouts. In these cases, we used the PSF-subtracted remainder images to measure position centroids and magnitudes.

\begin{figure}[ht]
\centering
\includegraphics[width=0.48\textwidth]{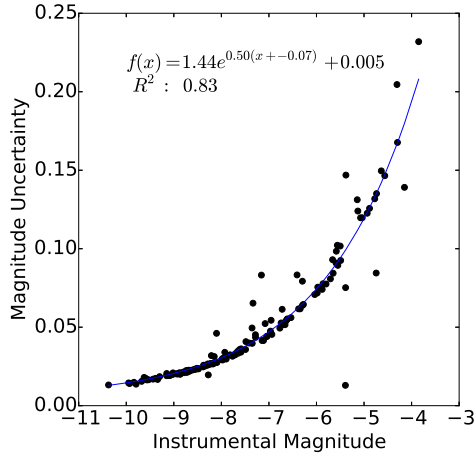}
\caption{
Photometric measurement error as a function of source brightness 
for the APT photometry. As expected, the uncertainties increase 
exponentially towards fainter magnitudes. The legend provides the
fit results and corresponding coefficient of determination, which ranges 
between 0-1 where 1 would indicate a perfect fit within the expected variance.
High outliers represent images with higher than typical noise. 
Low outliers occur in cases where secondary stars were found using 
the PSF-subtracted image.  An approximate zero point scaling is that -8 mag
on the instrumental scale corresponds to roughly 12.75 mag on a Vega scale.
The photometry thus spans the magnitude range $\sim$10.5-16.5 mag.
}
\label{fig:maguncert}
\end{figure}

We explored for each image both a constant aperture of radius of 5 pixels 
(0.11\arcsec, containing about 83\% of the encircled energy),   
and a custom aperture of radius 10-30 pixels, sufficient to cover 
92-96\% of the flux, even for the poor-quality images, 
for each member of each binary.  
The magnitude difference discrepancies between the constant aperture 
and the custom aperture form a roughly gaussian distribution. 
Our final photometry values come from the smaller aperture,
in order to avoid contamination from the other member in the pair.
Figure~\ref{fig:maguncert} illustrates the APT-reported measurement error
as a function of the APT-reported instrumental magnitude.  

For all detected binary systems, we computed magnitude differences 
and associated errors.  We performed an independent check of the photometry
using the pipeline described in Law et al. (2014) and Ziegler et al. (2017).
For sources with close separations, the later values are preferred given
the explicit de-blending, reducing the measured contrast for these objects
over straight aperture photometry.  Excluding these outliers, the mean contrast
difference between the two methods was 0.01 mag and the dispersion 0.14 mag,
with point-to-point agreement at the $<$1-2 sigma level.

A significance for each companion detection was calculated 
using the methods employed in Ziegler et al. (2017). Briefly,
the local noise as a function of separation from the target star is measured by sliding a 10-pixel diameter aperture within concentric annuli centered on the target star.  The signal within the aperture at each position is measured, and the mean and standard deviation of the set of signals in each annulus is calculated.  If the annulus contains an astrophysical source, the measured signals for that annulus are sigma clipped to remove the outlier signals associated with the source.   An aperture is subsequently placed on the observed nearby star, and the signal compared to the local noise to estimate the detection significance.

Table~\ref{tab:bin} reports contrast values 
and a significance for each companion detection,
both measured as described above.

\section{Analysis of Images with No Detected Companions: Contrast Limits}
\label{sec:analysis}

\begin{figure}[htpb]
\centering
\includegraphics[width=0.48\textwidth]{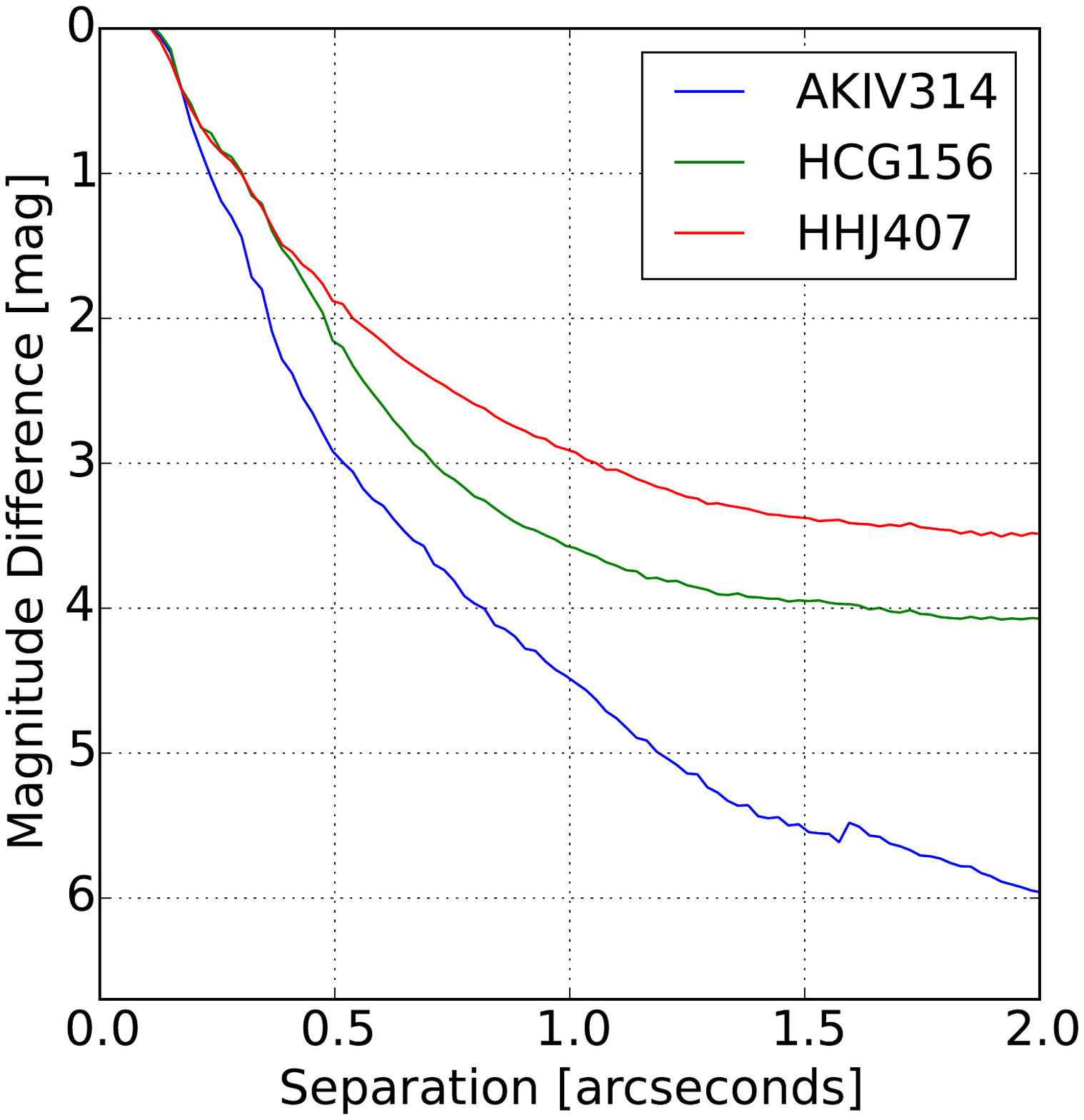}
\includegraphics[width=0.48\textwidth]{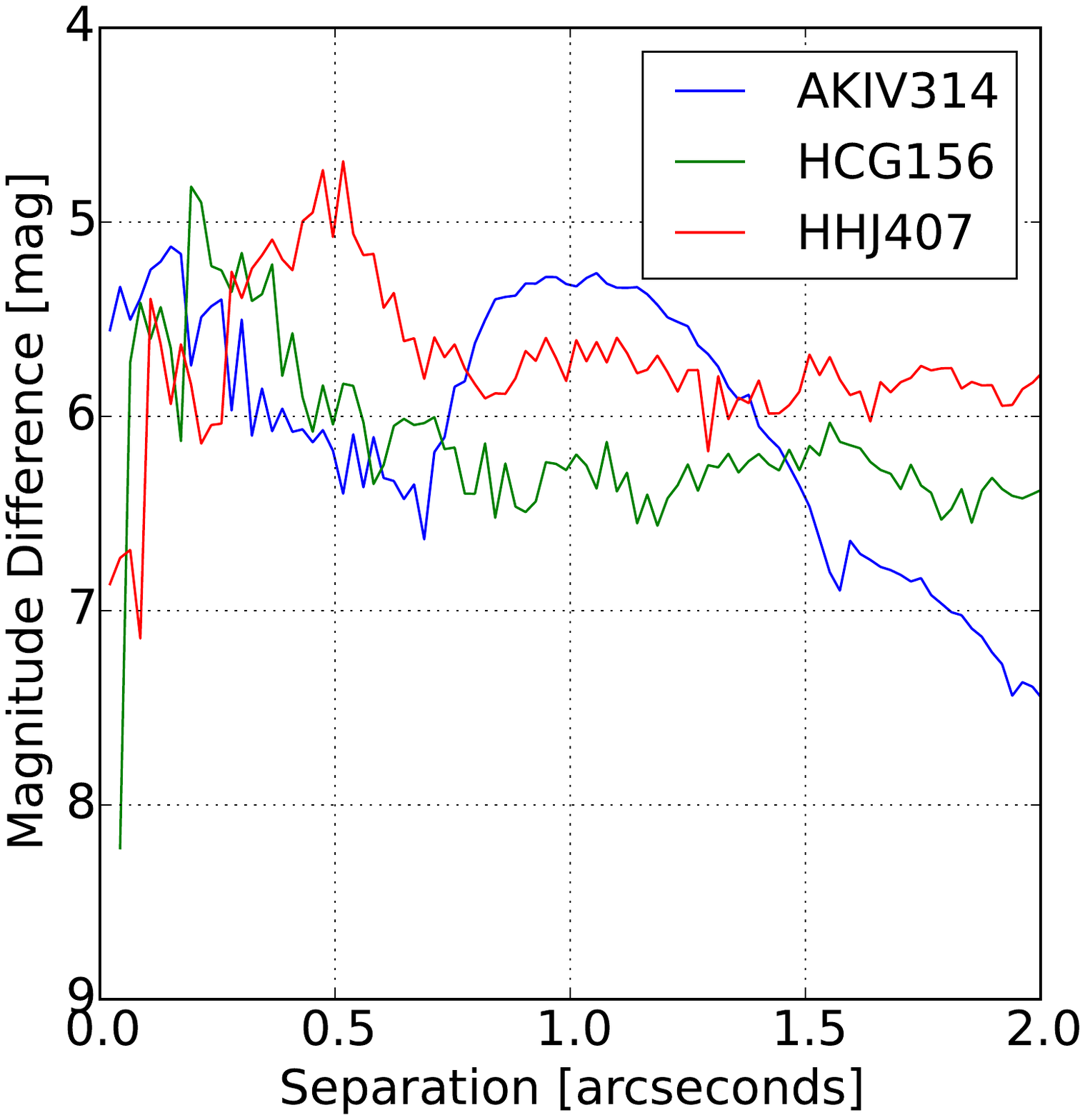}
\caption{
Contrast curves for three stars of varying image quality; raw contrast
is on the left and PSF-subtracted contrast on the right. 
The stars illustrated are: AK IV-314, which was one of the most well-imaged 
stars, but is a binary system with separation of 1.03\arcsec; the secondary 
has been subtracted before generating the contrast curves 
(resulting in the artifact discontinuity around 1.6\arcsec). 
HCG-156 is representative of the mean performance, and is a single star. 
HHJ-407 is about one standard deviation away from the mean performance. 
As illustrated in the right panel, PSF subtraction improves the contrast 
by several magnitudes, though below $\sim$0.2\arcsec\ 
(roughly $2\times \lambda/D$), companion detection 
is still not possible.  The broad bump for AK IV-314 is the residual 
of the seeing halo of the subtracted companion.
}
\label{fig:contrast}
\end{figure}

Only a fraction of the observed sources had identifiable companions.
To assess our ability to detect binary stars and determine upper limits
for undetected companions, we generated contrast curves for each final image.
A contrast curve denotes the separation-dependent relative brightness level a secondary star needs to exceed for 
detection in the image. 
Naturally, at decreasing separations from the primary, secondary stars need to be increasingly brighter in order to
be detected.  
An experimental contrast curve was generated by isolating the PSF of the
primary and background-subtracting, then finding the median counts within a sky annulus encompassing 38 to 43 pixels from each centroid position. 
The PSF was then scaled down and placed into the original image at a
series of random separation and random primary-secondary magnitude differences. The modified image was examined visually to see if the inserted secondary stars
could be detected by eye, leading to rough contrast estimates of 3-6$^m$ at 0.5-3\arcsec.

A more sophisticated algorithm for theoretical contrast curve generation 
involved first converting all pixels into polar coordinates with origin at the
primary star centroid. Next, around a circle describing the polar points at a given 
separation, we placed the maximum number of 5-pixel-radius tangent circles (apertures).
For separations less than 7 pixels, a set number of three such apertures were used, placed 120 degrees apart. For each of the 5-pixel-radius apertures around each circle,
the enclosed flux was measured, with the fluxes of partial pixels estimated using the ratio of areas
within and outside the 5-pixel-radius aperture. Then, the standard deviation and the mean of all 
the fluxes measured at a given separation was found. The $3-\sigma$ contrast limit was generated at each
separation using $3\times$ the standard deviation added to the mean flux level at that separation. 
Stated as a formula, we computed the contrast at separation $r$ as
$$C(r) = -2.5 \times log (SNR_{threshold} \times \sigma(S_r) + mean(S_r)) $$
where $S_r$ is the set of flux values within an annulus around $r$, and $SNR_{threshold}=3$.
For binary stars, the procedure was modified to exclude those 5-pixel radius apertures 
with centers located within 30-pixels of the secondary star center.

Contrast curves were generated for every final image.  
Figure~\ref{fig:contrast} illustrates the results for three examples of varying image quality. 
In Table~\ref{tab:sources}, we provide the contrast values as magnitude differences
at 0.5, 1, 2, and 3\arcsec\ angular separation, for all sources. 
The tabulated contrast curve data exhibit approximately normal 
distributions, with means and standard deviations at the four separations of:
$1.92\pm 0.46$ at 0.5\arcsec, $3.01\pm 0.63$ at 1\arcsec, $3.83\pm 0.84$ at 2\arcsec, and $3.97^m\pm 0.91^m$ at 3\arcsec.

In order to further investigate the data quality, we examined the relationship
between measured image FWHM and the change in the contrast values between 1\arcsec\
and 0.5\arcsec.  The correlation affirms that steep contrast curves come from
low FWHM, high quality data.  Anomalously poor contrast curves have large FWHMs
and are generally due to fainter sources and poor image quality. 

As a check on multiplicity, the contrast curves themselves were
examined for bumps that could indicate binary stars. One new binary,
of no particular dimness, that was overlooked in the earlier binary idenfication
analysis was found in this manner: CSIMon-0021.

\section{Cluster Multiplicity Results}
\label{sec:results}

Our Robo-AO survey covered about 10\% of the known cluster membership
in each cluster (ranging from 8\% of known NGC 2264 members to 14\% of known Pleiades members).
The NGC 2264 sample is comprised largely of FGK spectral types, while 
the Praesepe and Pleiades samples are largely K and early M types.

\begin{figure}[htpb]
\centering
\includegraphics[width=0.44\textwidth]{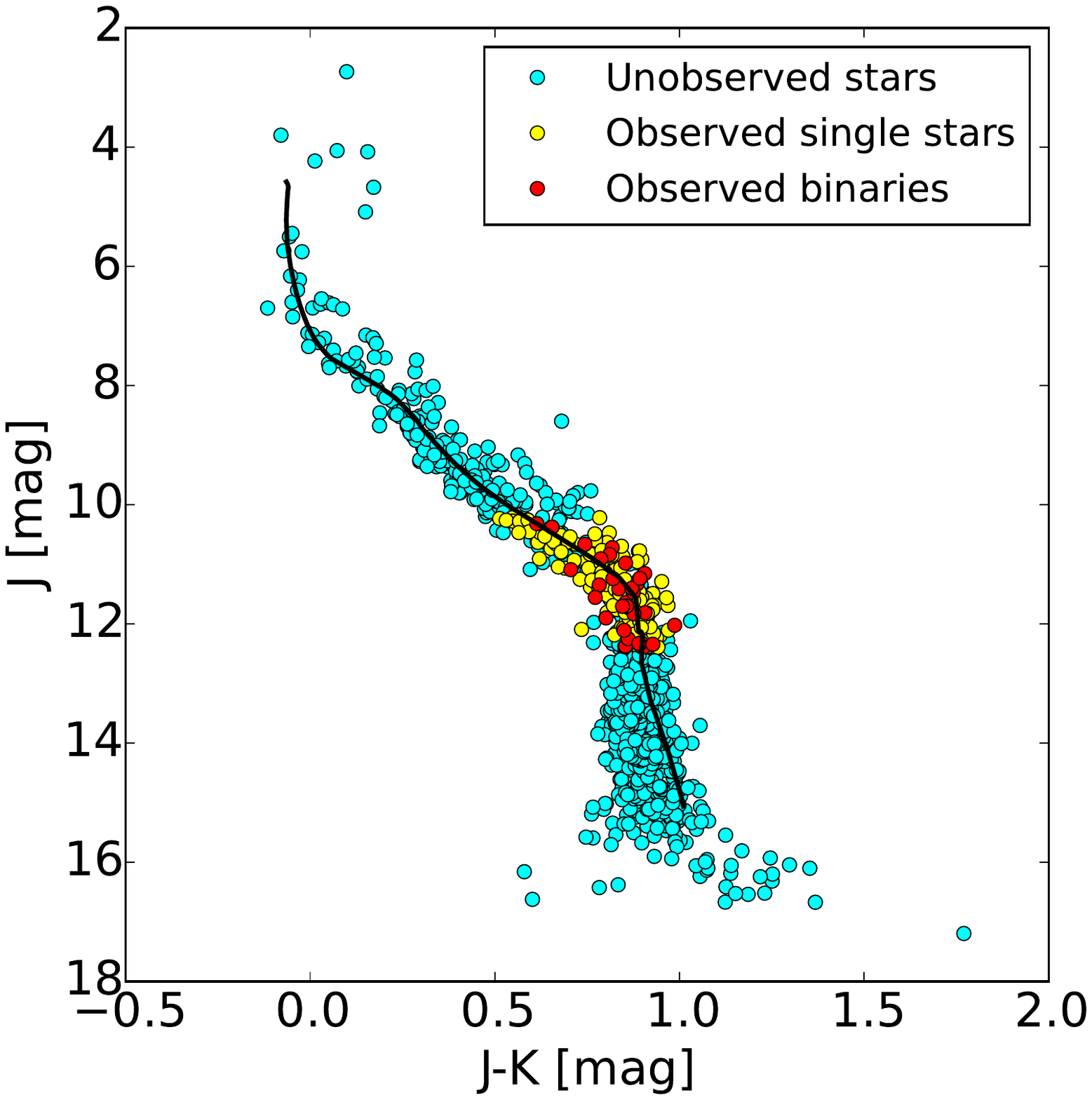}
\includegraphics[width=0.44\textwidth]{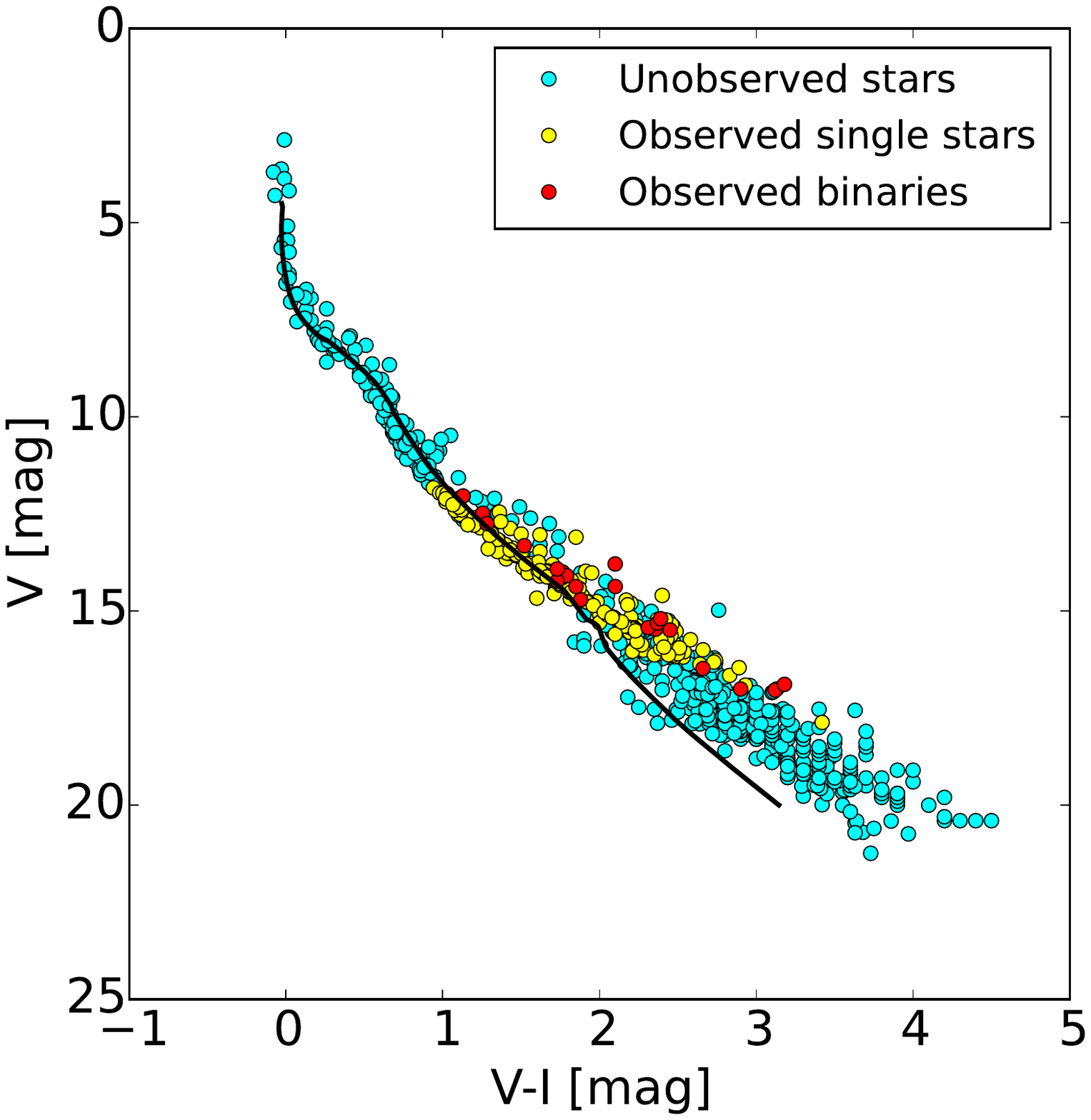}
\vskip-1.4truein
\includegraphics[width=0.44\textwidth]{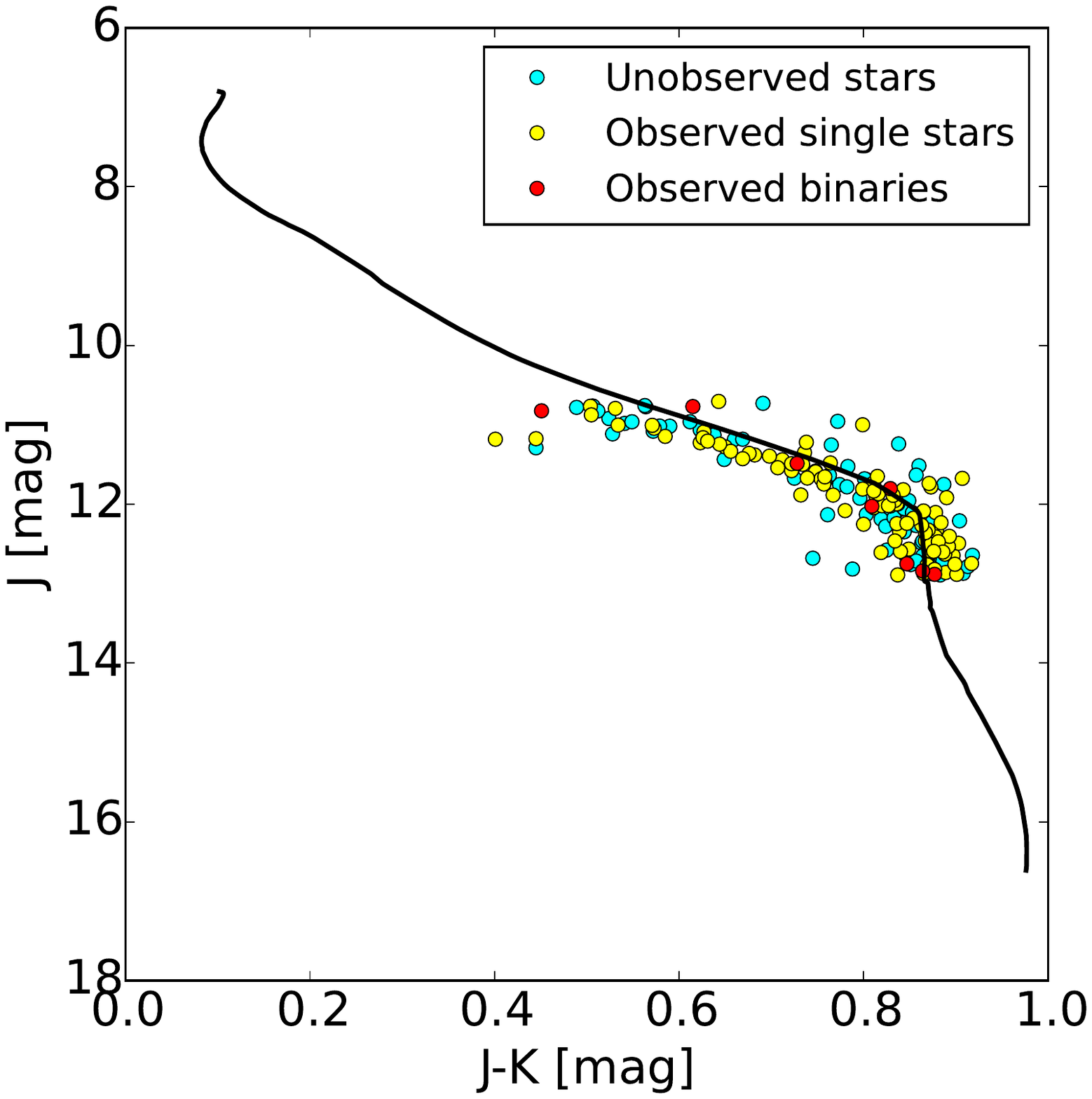}
\includegraphics[width=0.44\textwidth]{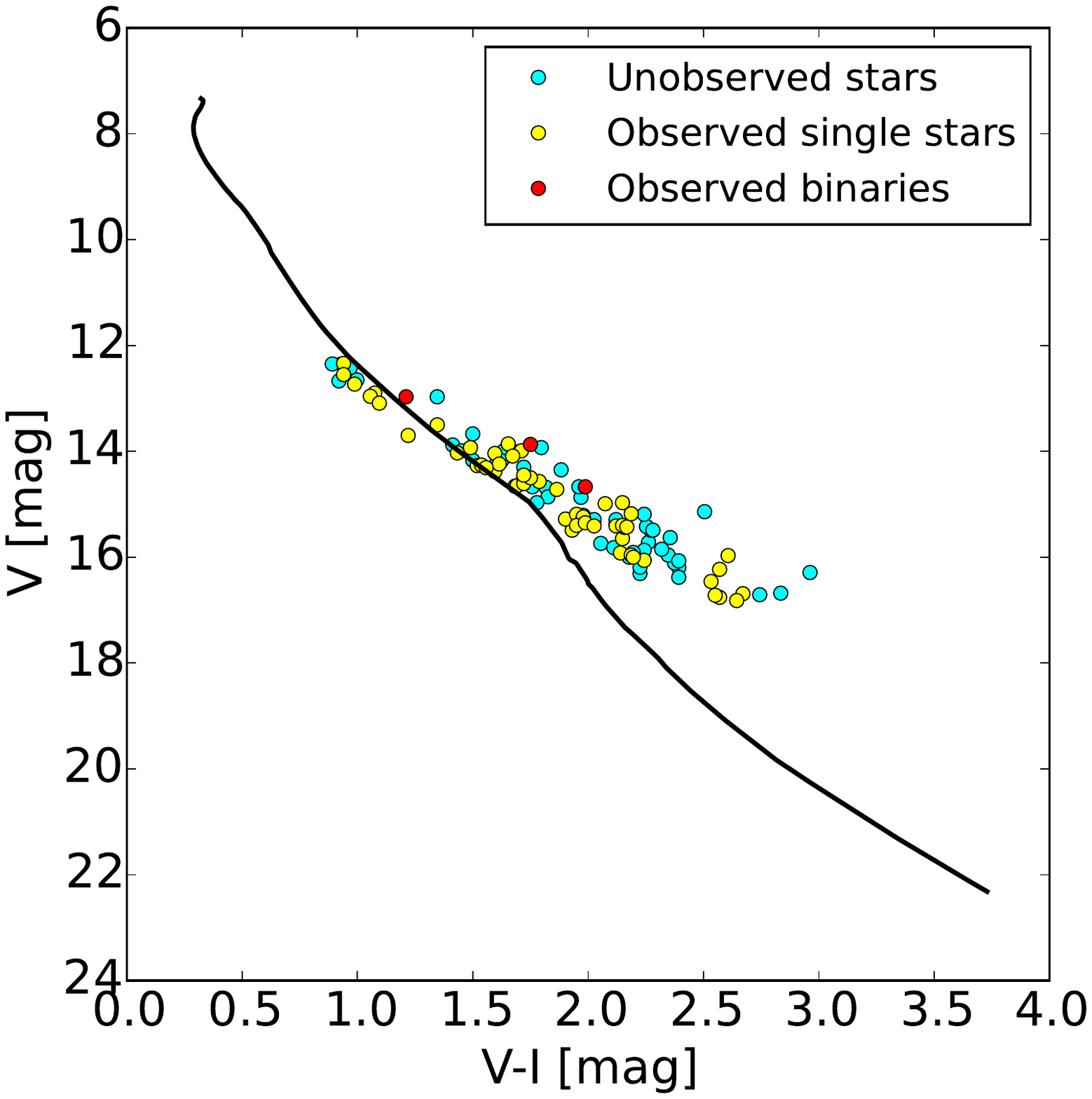}
\vskip-1.4truein
\includegraphics[width=0.44\textwidth]{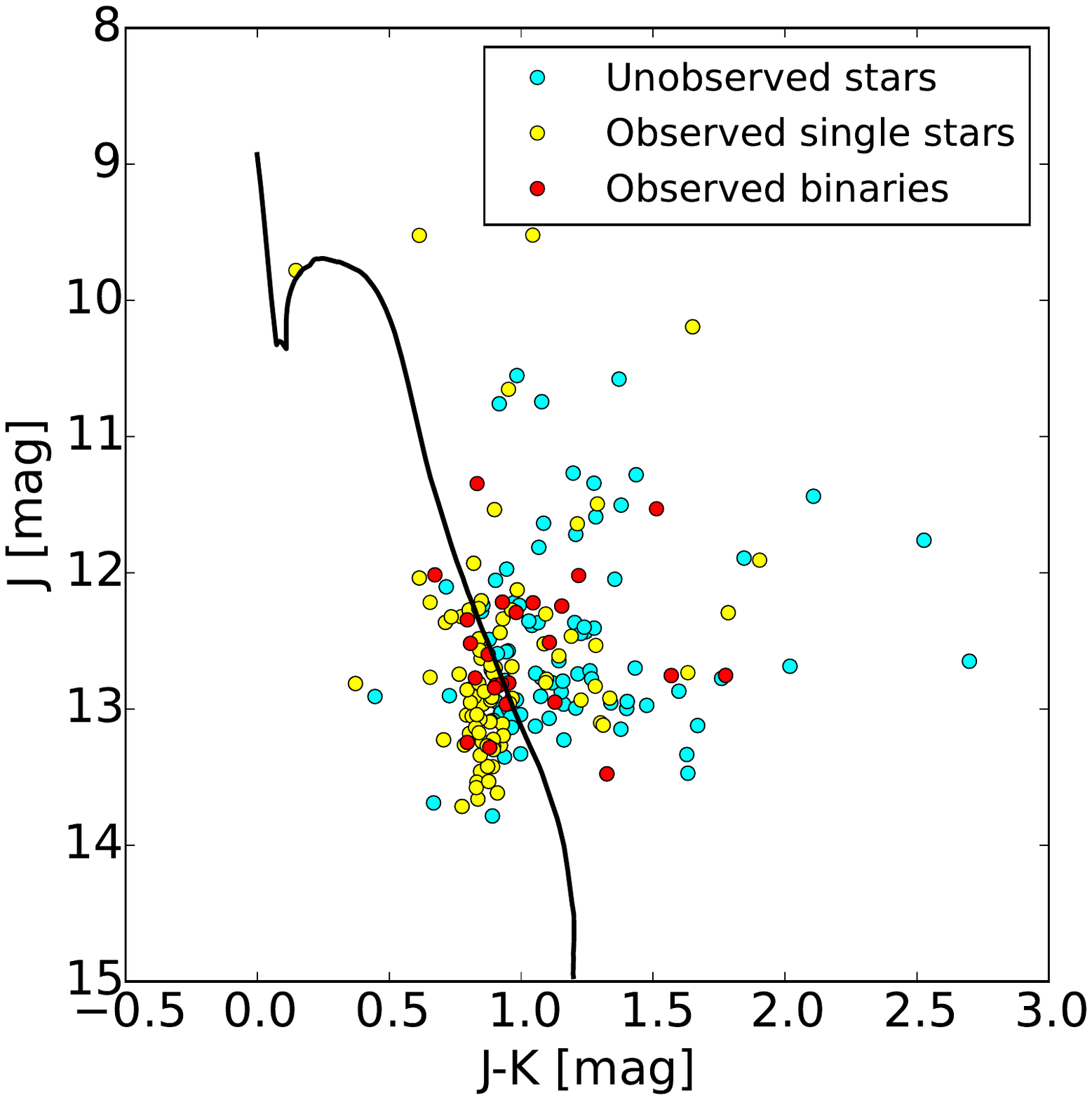}
\vskip-0.8truein
\caption{
Color-magnitude diagrams 
using 2MASS (Cutri et al. 2003) near-infrared and literature optical photometry
for the Pleiades (top), Praesepe (center), and NGC 2264 (bottom).
In each panel, shown are: unobserved cluster members for which
K2 time series photometry exists (cyan), Robo-AO observed single stars (yellow), 
Robo-AO detected pairs (red), and a theoretical isochrone at the
appropriate age, distance, and average reddening for the cluster.  
The Robo-AO observed samples typically span 
about two magnitudes in brightness within each cluster; for the Pleiades, the
full color-magnitude sequence is shown, for context, while for Praesepe and
NGC 2264 just the magnitude range of relevance to Robo-AO is shown. 
Clear cluster loci can be seen for Praesepe and the Pleiades,
but for NGC 2264 the locus is smeared due to the combined effects 
of circumstellar disks and extinction. 
}
\label{fig:cmds}
\end{figure}

\begin{figure}[htpb]
\centering
\includegraphics[width=0.44\textwidth]{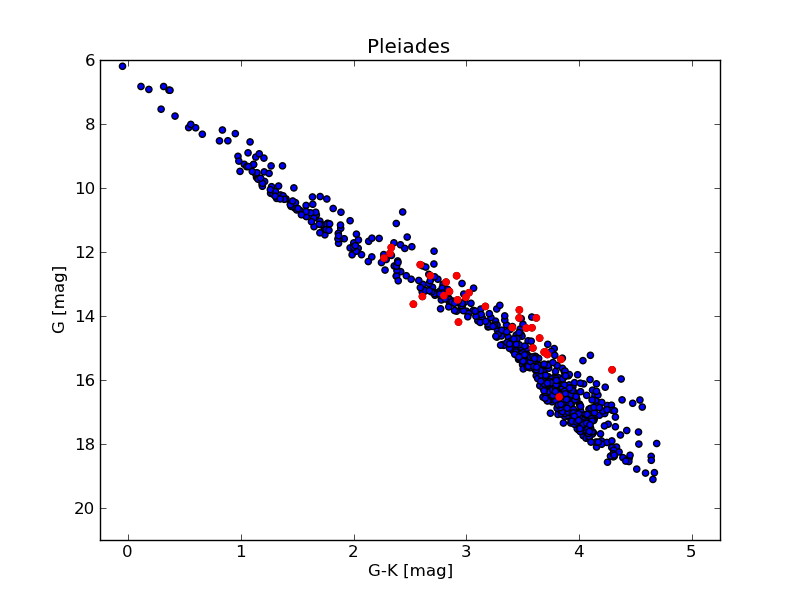}
\includegraphics[width=0.44\textwidth]{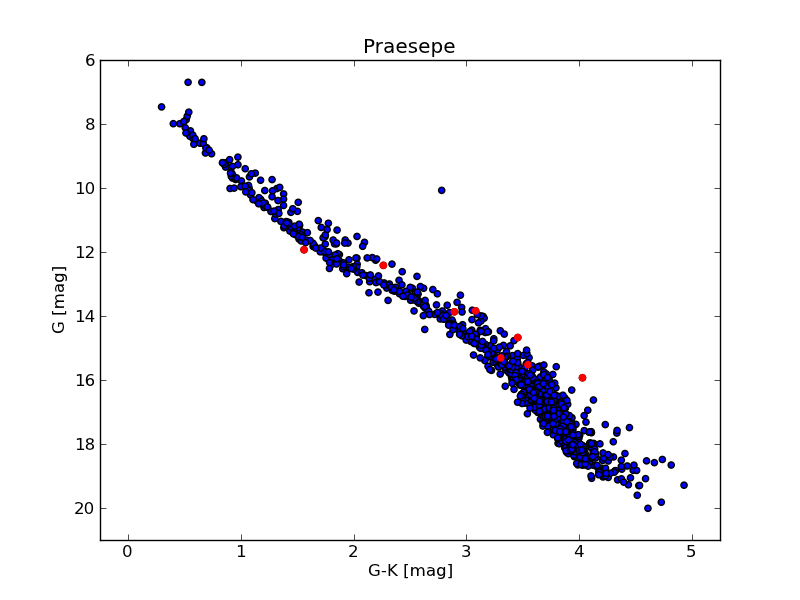}
\caption{
Location of the identified binaries (red) in an optical-infrared color-magnitude diagram,
based on Gaia G-band (van Leeuwen et al. 2017) and 2MASS K-band (Cutri et al. 2003) photometry of the entire clusters.  
Relative to the main Pleiades and Praesepe cluster sequences, identified binary pairs
stand out somewhere better in this color-magintude diagram than in 
the infrared-only color-magnitude diagram of Figure~\ref{fig:cmds}.
}
\label{fig:cmdsGK}
\end{figure}

Figure~\ref{fig:cmds} illustrates the color-magnitude diagrams for
known members of the three clusters, with those detected as singles or binaries
in Robo-AO data highlighted.
Figure~\ref{fig:cmdsGK} shows for
a broader sample of Pleiades and Praesepe members, 
a color-magnitude plane in which the binaries stand out 
from the main cluster sequences somewhat better.
The morphology of the Pleiades and Praesepe plots is consistent with the 
roughly main-sequence nature of these clusters, while the NGC 2264 plot
is both pre-main sequence and dominated by color excesses due to extinction
and dust disks.  The main point is that the distributions of both 
the Robo-AO observed stars, and the detected binaries,
are consistent with the general distribution of similar-color and magnitude objects.

We present a total of 66 candidate wide separation binaries - 32, 8, and 26 
in the Pleiades, Praesepe, and NGC 2264, respectively 
(see Table~\ref{tab:bin}).  Relative to the number of binaries indicated
in Table~\ref{tab:sources}, there are fewer sources listed in 
Table~\ref{tab:bin} since some members of pairs were found twice: 
as companions to each other when imaged separately.
Among the binaries, just 5, 2, and 0 in the three respective clusters
were revealed in a literature search as having their companions previously known. 
These confirmations (10\% of our reported binary sample)
provide confidence in our methods and detections overall, and further, allow 
the consideration of color information since the previously identified binaries 
had been observed at infrared wavelengths.
We expect the separations to be about the same
given the large cluster distances, but our reported magnitude differences 
will be larger in the optical than in the infrared, due to the expected 
redder color of the fainter companions, relative to their primaries.

\begin{figure}[htpb]
\centering
\includegraphics[width=0.95\textwidth]{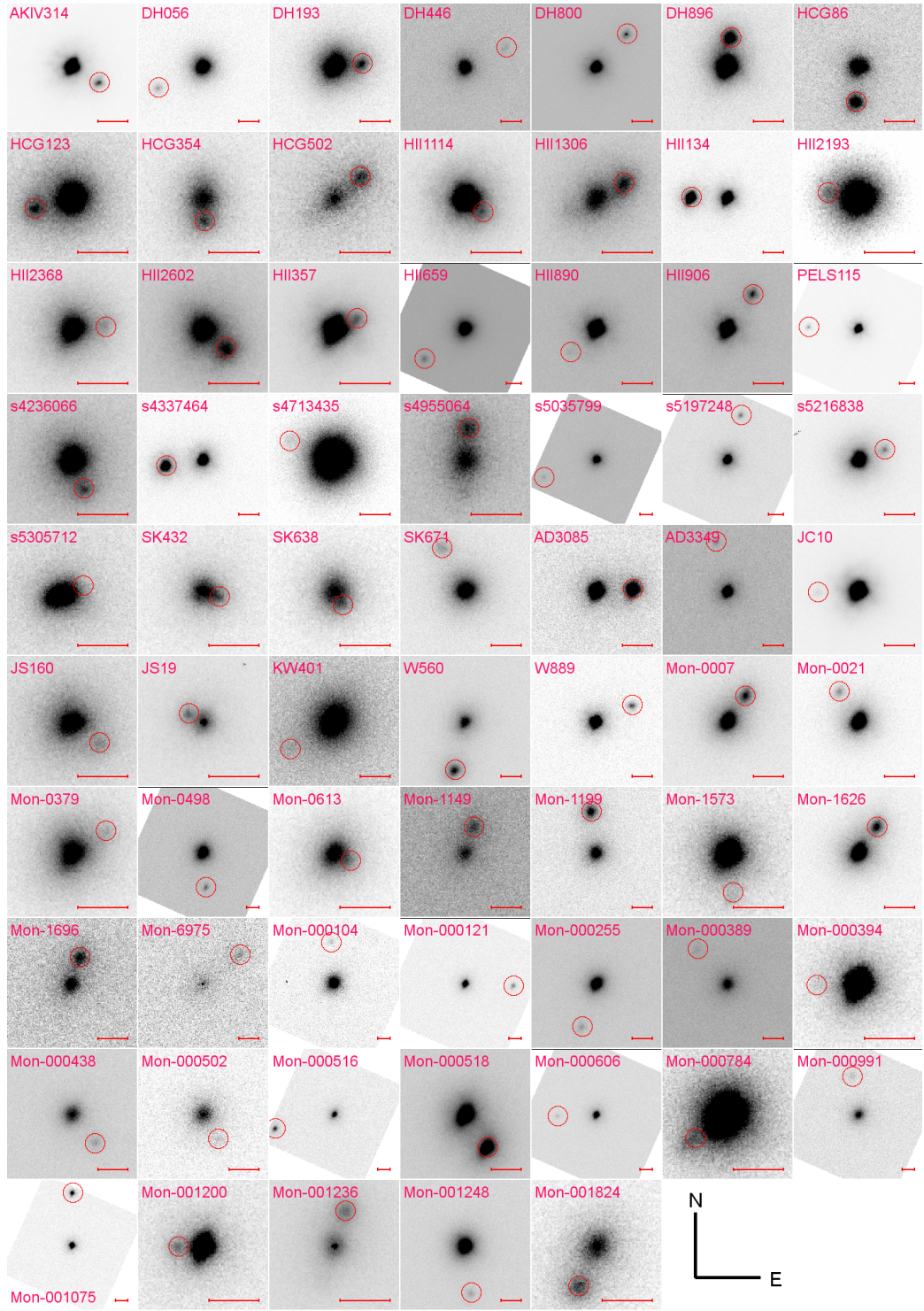}
\caption{
Robo-AO images of detected binaries; north is up and east is {\it to the right}.
Red circles indicate the secondaries and red bars indicate a constant angular size of 1\arcsec.
There are five different spatial scales, depending on the pair separation. 
Images are individually scaled in depth so as to highlight the binaries.
Labels correspond to lines in Table~\ref{tab:bin}, which provides the
separations, position angles, differential magnitudes in the red optical,
and detection significance for each system.
}
\label{fig:tiling}
\end{figure}

\begin{figure}[htpb]
\centering
\includegraphics[width=0.49\textwidth]{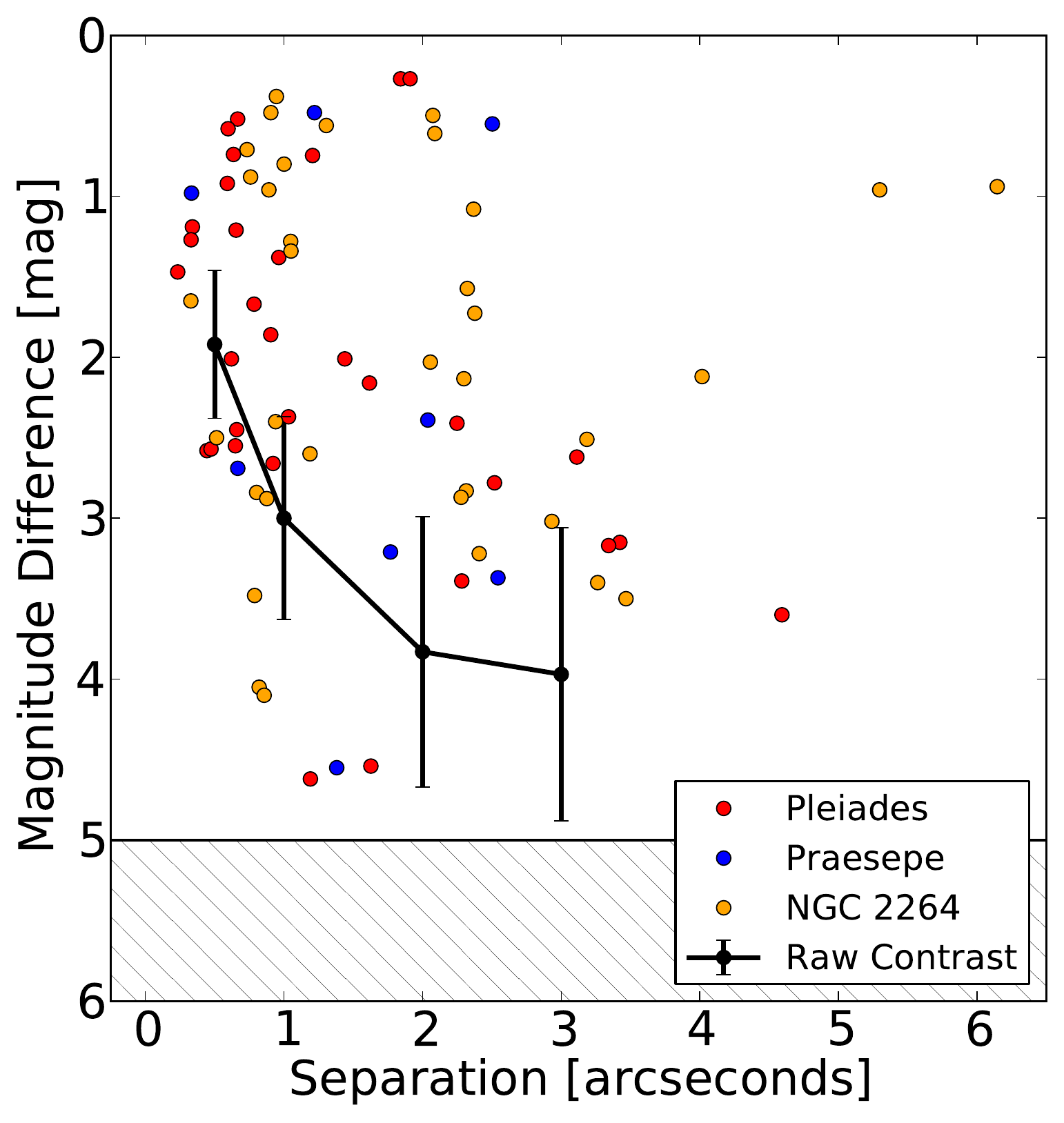}
\includegraphics[width=0.49\textwidth]{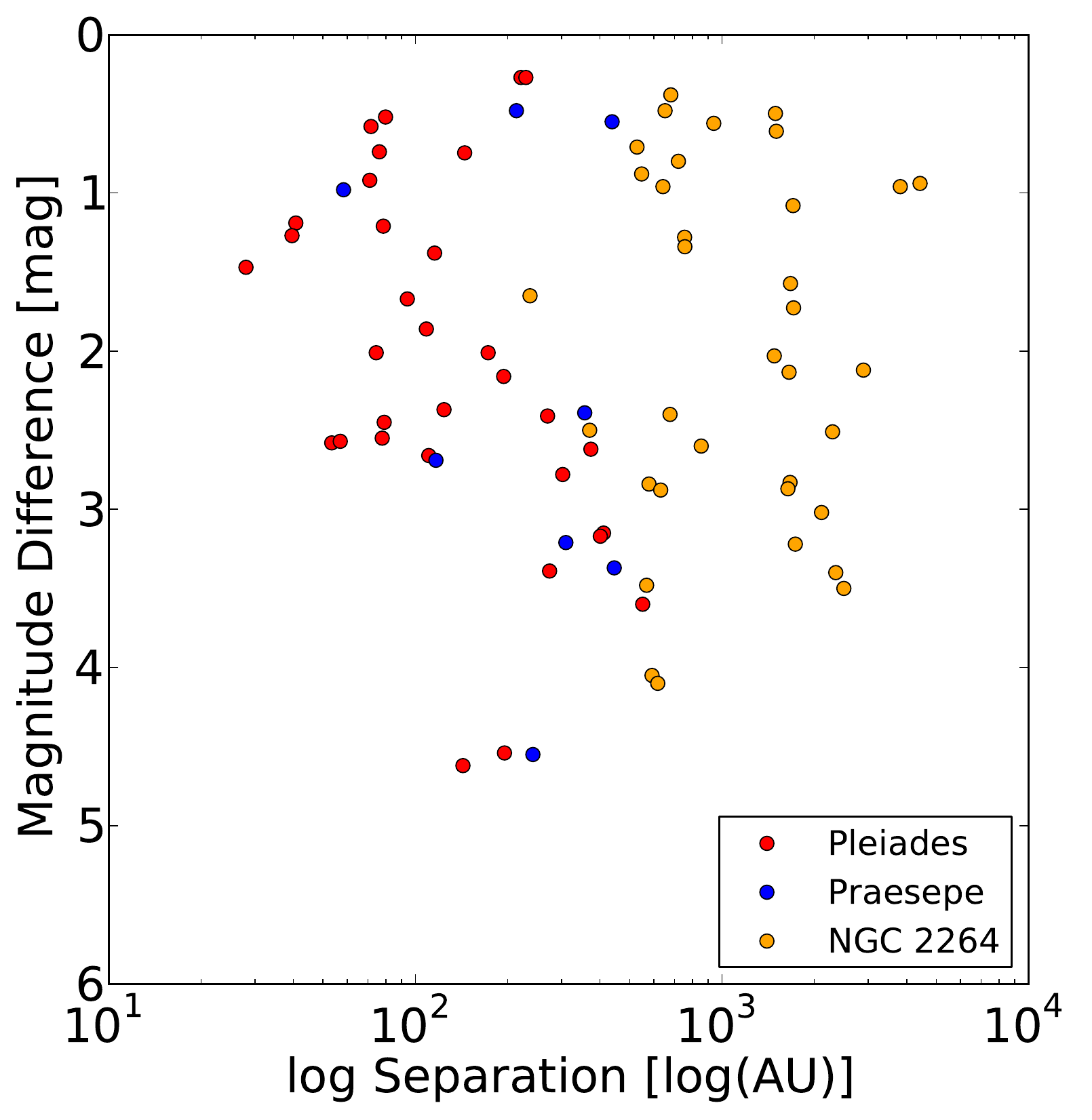}
\caption{
Separation vs. difference in magnitude for detected binary systems. 
Members of the three clusters are individually color-coded. 
Left panel is in empirical units, 
and includes the mean $3\sigma$ raw contrast curve,
and standard deviation among the contrast curves for the entire data set; 
PSF subtraction enables sensitivity 
down to 5-6 mag of contrast for most targets,  
as shown in Figure~\ref{fig:contrast}. 
Right panel is in physical units of AU; 
apparent is the reduced sensitivity to small-separation companions
for NGC 2264, due to the larger distance of this cluster,
as well as  the lack of sensitivity for Pleaides beyond $\sim$585-820 AU 
and Praesepe beyond $\sim$750-1050 AU, due to the maximum search radius
imposed by image size.
There is no correlation between separation and magnitude difference, 
in either panel.
}
\label{fig:sepmag}
\end{figure}

Figure~\ref{fig:tiling} presents the image gallery of our Robo-AO detected binary systems. 
There were 78 images containing candidate binaries, with 
70 unique sources excluding repeats and poor quality images.
Figure~\ref{fig:sepmag} illustrates the magnitude differences as a function of pair separation. 
Sources detected at higher contrast than the nominal limits come from 
better-than-average quality images or from the PSF-subtracted images;  
we note that this subset of high contrast companions is located 
within $\sim$1-1.5\arcsec, suggesting that they are true bound companions.  
The median pair separation is 1.1\arcsec\ with the 
peak between $\sim$0.5-1.0\arcsec\ and continued decline towards a flat 
distribution beyond $\sim$2\arcsec.  Beyond 4\arcsec, the data become 
incomplete in position angle due to the square images. In physical units,
the right panel of Figure~\ref{fig:sepmag} illustrates the rough segregation
of companion sensitivity by distance.
As NGC 2264 is further away, we were not sensitive to separations below 
$\sim$200-500 AU. Meanwhile, we detected binaries in the 30-500 AU range in
the Pleiades and Praesepe samples, and to higher contrast levels than in
NGC 2264. 

Notably, there are many images on which binaries are detected at
magnitude differences exceeding the individual 
``raw" contrast values listed in Table~\ref{tab:sources},
and/or are more than three standard deviations better than the nominal contrast 
at a given separation (left panel of Figure~\ref{fig:contrast}).
These high contrast outliers are due to the use of PSF-subtracted images, 
which were generated for all final images.  The PSF-subtracted data
naturally have significantly better contrast, down to below 5 mag 
for most targets, as illustrated in the right panel for the three examples 
shown in Figure~\ref{fig:contrast}.
Five binary systems have brightness ratios of more than 
four magnitudes.  Two of these, including the most significant outlier 
at the smallest separations, could not be seen in the initial image cutouts, 
but are clearly present in the PSF-subtracted images 
(and were detected in that manner). The other high brightness ratio systems
were identified in the non-PSF-subtracted data in observations with
exceptionally good contrast curves.

Results for the individual clusters are presented below.
The raw multiplicity fraction across the three clusters, counting a
total of 70 visual binary systems among 441 distinctly imaged targets 
with good Robo-AO imaging, is thus 
15.9\% $\pm$ 1.9\% (or 15.1\% $\pm$ 1.9\%    
if duplicate pairings where each component was observed as the primary 
are removed).  This fraction is comparable to the fraction of multiples found 
by Baranec et al. (2016) and Ziegler et al. (2017) using the same equipment but 
targetting a much older field star sample of comparably bright sources.  
However, the binary fraction varies significantly 
among the clusters with NGC 2264 (where we are sensitive to mainly wider
separations and smaller magnitude differences, in a younger cluster) 
$\sim$50\% higher, Pleiades at about the mean value, 
and Praesepe $\sim$50\% lower.   
Because of the different cluster distances and ages, 
as well as the varying observing conditions during data acquisition,
the sensitivity to companion separation and mass varies.  Furthermore, the 
targets have different primary masses in NGC 2264 relative to the 
Pleiades and Praesepe samples, since the observations were constrained within
the same apparent brightness ratio.  For these reasons, the
necessary but complex and uncertain incompleteness corrections 
required to turn our raw multiplicity fractions into fractions for specific
$a$ and $q=m_2/m_1$ ranges, given the primary star $m_1$ ranges, have not been applied.

\subsection{The Pleiades at 136 pc}

In the Pleiades cluster, 34 binary systems (including 2 pairs that were observed twice and identified
as binaries of each other, making 32 unique pairs) were found among 212 targets. 

HII 1306, a 0.60\arcsec\ separation and $\Delta m_{opt}$=0.58$^m$ system was also identified
as a binary by Richichi et al. (2012), who reported separation 0.65\arcsec\ and $\Delta m_K$=0.33$^m$. 
HII 134, a 1.84\arcsec\ separation and $\Delta m_{opt}$=0.27$^m$ system was also identified
as a binary by Bouvier et al. (1997), who reported separation 1.83\arcsec\ and $\Delta m_K$=0.13$^m$. 
Bouvier et al. (1997) also found HII 2193 which we identify as
a separation 0.66\arcsec\ and $\Delta m_{opt}$=2.45$^m$ system; 
the previously reported separation is 0.69\arcsec\ and $\Delta m_K$=1.71$^m$. 
Finally, for HII 357 we report a companion with separation 0.47\arcsec\ and $\Delta m_{opt}$= 2.57$^m$ 
while Bouvier et al. (1997) found separation 0.50\arcsec\ and $\Delta m_K$=1.64$^m$. 

For HII 890 we report a companion at separation 1.19\arcsec\ and $\Delta m$=4.62$^m$ 
while Bouvier et al. (1997) found separation 1.74\arcsec\ and $\Delta m_K$=5.41$^m$, 
declaring the secondary source a background field object.
This is the only detected pair with apparent relative motion between the components, 
and it is also the only pair with a bluer color for the secondary relative to its primary.

A raw multiplicity of 16.0\% $\pm$ 1.4\% was
obtained from our observations of Pleiades KM primaries.

{\it Notes on individual sources: }
One of the final images, HII 370, contained a false tripling artifact of the image acquisition, 
and thus could not be analyzed. One of the single stars, s4798986, was also excluded from analysis due to poor
imaging. Three likely single stars, BPL 273, HII 34, and HCG 194, were ``lumpy" rather than cleanly detected
as binaries.  Uncertainties regarding these targets add to the uncertainty of our reported multiplicity fraction.

\subsection{Praesepe at 175 pc}

Out of the 108 Praesepe targets, 8 were observed to be likely binaries. 

As for the Pleiades, several of our optically identified Praesepe systems 
appear in previous literature announcing detected companions.  
KW 401 is reported as a 1.77\arcsec\ separation and $\Delta m_{opt}$=3.21$^m$ system 
while Bouvier et al  (2001) detected KW 401 as a triple system 
with one component at 1.69\arcsec\ and $\Delta m_K$=2.30$^m$ at a similar position angle
as our detection, and another component at separation 1.78\arcsec\ and $\Delta m_K$=5.4$^m$ 
that we do not see, presumably due to large optical-infrared color. 
We also detected W 560 as a 2.50\arcsec\ separation $\Delta m_{opt}$=0.55$^m$ system which 
Bouvier et al  (2001) report with separation 2.43\arcsec\ and $\Delta m_K$=0.92$^m$.

The observed multiplicity frequency for our Praesepe sample of KM primaries,
is 10.2\% $\pm$ 1.9\%.

{\it Notes on individual sources: }
One of the binary systems, JS 494, was excluded from analysis due to an additional false tripling artifact of the image acquisition. 
JS 231, a likely single star, and KW 566, likely a binary system, both have poor image quality and also
could not be properly analyzed.
Uncertainties regarding these targets add to the uncertainty of our reported multiplicity fraction.

\subsection{NGC 2264 at 740 pc}

Among 120 NGC 2264 targets, 34 (including 8 pairs that were observed twice and identified
as binaries of each other, making 26 unique pairs) likely binaries were detected. 
There is no previous dedicated study of visual binaries in this cluster;
a few such sources are known within the separation range 
to which we are sensitive (e.g. S Mon at 27 AU, R Mon at 530AU, and
AR6 at 2100 AU), but we did not observe any of these objects. 
The observed multiplicity for our NGC 2264 target group of FGK primaries, is 27.3\% $\pm$ 4.1\%.

{\it Notes on individual sources: }
Images for likely binary systems CSIMon-0394, CSIMon-0890, and CSIMon-0894 were poor, and thus photometry
could not be performed so the 3 targets were excluded in our analysis. CSIMon-0618 and
CSIMon-0486 are most likely singles, but also are of poor quality. 
Uncertainties regarding these targets add to the uncertainty of our reported multiplicity fraction.

\subsection{Mass Ratios}

The mass ratio of each identified binary system was estimated using theoretical isochrones that relate
magnitude to mass. The pre-main sequence and main sequence isochrones 
of Siess et al. (2000) were employed in conjunction with
the NextGen+AMES atmospheres of 
Haushildt, Allard, \& Baron, (1999) and Allard et al. (2000) to generate V, I$_C$, J, and K$_s$ magnitudes.  We interpolated isochrones at 125 Myr and 
$A_V=0.15$ mag, assuming a DM=5.67 for the Pleiades, 
and at 757 Myr and $A_V=0.00$ mag, assuming a DM=6.22 for Praesepe.
Due to the significant and target-dependent extinction in NGC 2264,
combined with the potential influence of circumstellar disks at J and K, 
the mass ratio exercise was not carried out for members of this star-forming region.
Color-magnitude diagrams for the Pleiades and Praesepe clusters in J vs J-K$_s$,
J vs J-H, and V vs V-I$_C$ demonstrate that the chosen isochrone set is a reasonable match 
to the cluster sequences.  Generally, calculated isochrones are too blue and/or too faint for lower mass stars, especially in V vs V-I$_c$, compared to open cluster data; 
the Siess et al. (2000) models are a closer match to empirical data than most.
We adopt J vs J-K$_s$ for the mass decomposition, due to both the isochrone match
and the uniform availability of data from 2MASS for our sample stars.

Previously measured magnitudes of our targets consist of the combined brightness
of both stars for our identified pairs.  These composite magnitudes 
(Table~\ref{tab:sources}) were decomposed using the flux ratios tabulated in Table~\ref{tab:bin}. 
While the flux ratios are measured in the LP600 or the SDSS i' filters,
and would roughly correspond to previous measurements in the I$_C$-band,
such I$_C$-band magnitudes are not readily available for most of the sample.
Instead, the abundant J and K$_s$ magnitudes from 2MASS were used.
Furthermore, as illustrated in Figure~\ref{fig:cmds}, the near-infrared
brightness and color predictions are better in the near-infrared than in
the optical for low mass stars.
For each cluster, the point on the age-appropriate 
J vs J-K$_s$ isochrone closest to each binary system was found. 
Then the theoretical $I_C$ magnitude corresponding
to these J and K$_s$ magnitudes was decomposed into the constituent magnitudes
of the two stars. Once again using the isochrones,
the corresponding mass to each magnitude was found, and a ratio
was determined for primary star $m_1$ and secondary star $m_2$.
As the isochrones did not include stars under 0.1 $M_\odot$,
for some of the faintest secondaries only an upper limit
for the mass ratio was obtained.

\begin{figure}[htpb]
\centering
\includegraphics[width=0.48\textwidth]{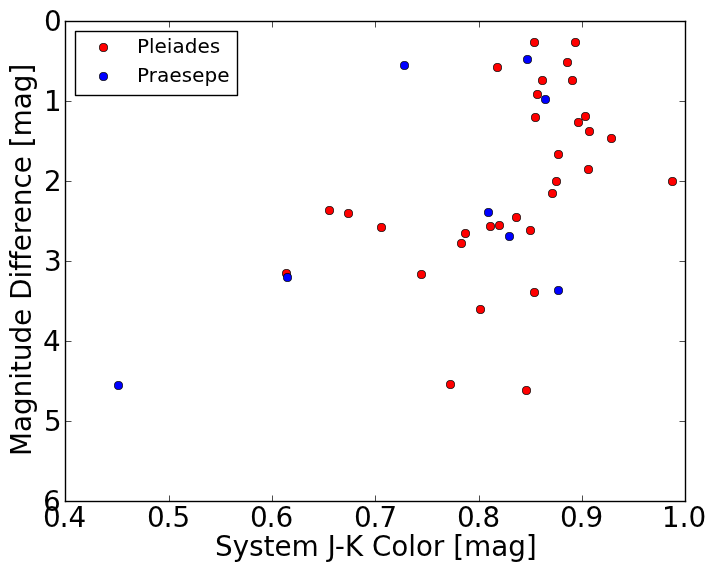}
\includegraphics[width=0.48\textwidth]{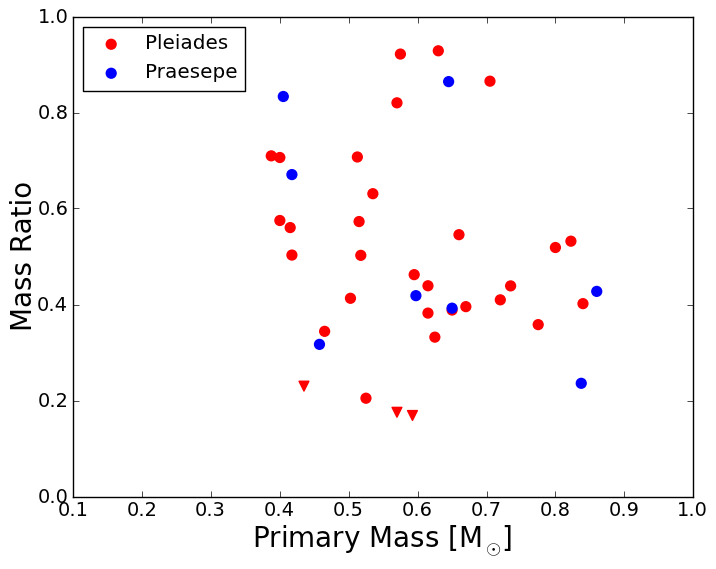}
\caption{
Magnitude difference as a function of system composite color (left) and
corresponding inferred mass ratio as a function of primary mass (right)
for the Pleiades (red, with upper limits indicated by inverted triangles
set by the lowest masses available from the adopted evolutionary tracks) 
and Praesepe (blue) binary samples. 
As the distributions are dominated by small number statistics, no conclusions
can be drawn beyond illustration of the range in $q$ of the detected binaries. 
}
\label{fig:massratios}
\end{figure}

Figure~\ref{fig:massratios} illustrates the mass ratios as a function of
primary mass.
The Robo-AO data set spans a range in $q = m_2/m_1 \approx 0.2-0.9$
(see Table~\ref{tab:bin}). 
Small number statistics, and the lack of incompleteness corrections, prevent
us from drawing any conclusions regarding the true mass ratio distribution
at the wide separations probed by the data set (Figure~\ref{fig:sepmag}).

\section{Discussion and Summary}
\label{sec:summary}

Our optical multiplicity survey of 
120 members of NGC 2264,  
212 Pleiades members,   
and 108 Praesepe members  
covered in each case approximately 10\% of the cluster membership 
cataloged to date.  In NGC 2264, ours is the first high spatial resolution survey. 
In the Pleiades and Praesepe, previous similar work 
(notably by Bouvier et al. 1997 in the Pleiades, and by Bouvier et al. 2001 and 
Patience et al. 2002 in Praesepe) has been conducted at infrared wavelengths,
though in sum covering a comparable number of stars to our study. 

We identified 66 unique binary systems, only 7 of which were previously known. 
Given the small contrast ratios and close separations,
the majority of the newly identified companions are likely to be physically bound.
Towards the low-contrast widest pairs, and the highest contrast close pairs, boundedness becomes less likely.
We can assess the likelihood of chance superposition of faint foreground or background objects using a simulation of galactic stellar populations.  
We queried the TRILEGAL (Girardi et al., 2005, Girardi et al. 2012) 
V1.6 model\footnote{http://stev.oapd.inaf.it/cgi-bin/trilegal} 
over 1 deg$^2$ fields towards each of the clusters, so as to establish a
representative sampling of the contamination on few arcsec scales.  
The raw star counts 
brighter than $i=17$ mag (see Figure~\ref{fig:maguncert})
were scaled down to the 8.6\arcsec\ by 8.6\arcsec\ field-of-view of RoboAO 
to arrive at an expected 0.0025, 0.0181, 0.0087 contaminating stars per 
observation towards the Pleiades, Praesepe, and NGC 2264 clusters, respectively.
Multiplying by the number of sources observed in each cluster,
results in 0.5, 2.0, and 1.0 expected interlopers that we could be incorrectly 
calling binaries.  However, these numbers are upper limits if one considers
the smaller separations actually occupied by the observed companion 
distribution (see left panel Figure~\ref{fig:sepmag}).  Specifically, we
would expect only 0.08, 0.33, and 0.17 total contaminants from the three
clusters at $<3.5$\arcsec\ -- where nearly all of the observed companions are located -- 
and that any such contaminants would be close to the magnitude limit and
therefore at the higher contrast levels.
Future work involving proper motions and colors is required in order to 
definitively establish binarity versus field star contamination.

Our observations were sensitive to only those companions located
beyond the peak of the separation distribution produced by 
Duquennoy \& Mayor (1991) and
Raghavan et al. (2010) for solar neighborhood field stars.  
Our observations also targetted only a narrow magnitude range, which
corresponds to different primary mass ranges and secondary mass sensitivities
at the different cluster distances.
Thus, we can not make meaningful comparisons to the features of the
field star distributions in either separation or mass ratio.

Nevertheless, the results of our work broadly 
sample mass ratios $q=m_2/m_1=0.2-0.9$ 
around primary stars with $\sim0.4-0.9M_\odot$.
The measured parameters for individual objects will be valuable 
for future studies aimed at placing the multiplicity results 
for single-age clusters in the context of field star samples
which have diverse ages, though better characterized multiplicity properties.
Additional high spatial resolution survey work that would complete
our multiplicity census for these important clusters should be carried out.

\acknowledgments
We acknowledge the Caltech SURF (Summer Undergraduate Research Fellowship) 
program for partially supporting the work of C. Zhang.
Russ Laher provided appreciated guidance on use of the APT.
Luisa Rebull provided the catalog photometry used in creating Figure~\ref{fig:cmdsGK}.
We thank Emma Hovanec for literature research on the targets of this study.
The Robo-AO system at Palomar was supported by the collaborating partner institutions, the California Institute of Technology and the Inter-University Centre for Astronomy and Astrophysics, and by the National Science Foundation under grants AST-0906060 and AST-0960343, by the Mount Cuba Astronomical Foundation, and by a gift from Samuel Oschin. 
C.B. acknowledges a Sloan Fellowship.
We remain grateful to the Palomar Observatory staff for their efforts in support of astronomical research,
and to the Palomar Observatory Director for the allocation of observing time to this project.

\bibliography{main}
\section{References}

Adams, J.~D., Stauffer, J.~R., Skrutskie, M.~F., et al.\ 2002, \aj, 124, 1570 

Affer, L., Micela, G., Favata, F., Flaccomio, E., \& Bouvier, J.\ 2013, \mnras, 430, 1433 

Allard, F., Hauschildt, P.~H., \& Schweitzer, A.\ 2000, \apj, 539, 366

Baranec, C., Riddle, R., Law, N.~M., et al.\ 2014, \apjl, 790, L8 

Baranec, C., Ziegler, C., Law, N.~M., et al.\ 2016, \aj, 152, 18 

Baxter, E.~J., Covey, K.~R., Muench, A.~A., et al.\ 2009, \aj, 138, 963

Bettis 1975, PASP 87, 797   

Bodenheimer, P.~H.\ 2011, Principles of Star Formation: , Astronomy and Astrophysics Library.
Springer-Verlag Berlin Heidelberg, 2011 (Chapter 6).

Bolte, M.\ 1991, \apj, 376, 514 

Boss, A.~P.\ 1995, Revista Mexicana de Astronomia y Astrofisica Conference Series, 1, 165

Boudreault, S., Lodieu, N., Deacon, N.~R., \& Hambly, N.~C.\ 2012, \mnras, 426, 3419

Bouvier, J., Rigaut, F., Nadeau, D., 1997, Astronomy and Astrophysics, 323, 139.

Bouvier, J., Duch{\^e}ne, G., Mermilliod, J.-C., \& Simon, T.\ 2001, \aap, 375, 989 

Bouy, H., Moraux, E., Bouvier, J., et al.\ 2006, \apj, 637, 1056 

Bouy, H., Bertin, E., Sarro, L.~M., et al.\ 2015, \aap, 577, A148

Cody, A.~M., Stauffer, J., Baglin, A., et al.\ 2014, \aj, 147, 82

Cutri, R.~M., Skrutskie, M.~F., van Dyk, S., et al.\ 2003, The IRSA 2MASS All-Sky Point Source Catalog, NASA/IPAC Infrared Science Archive.


Duch{\^e}ne, G., \& Kraus, A.\ 2013, \araa, 51, 269 

Duquennoy, A., \& Mayor, M.\ 1991, \aap, 248, 485

Garcia, E.~V., Dupuy, T.~J., Allers, K.~N., Liu, M.~C., \& Deacon, N.~R.\ 2015, \apj, 804, 65 

Gillen, E., 2015, Ph.D. Thesis, University of Oxford

Gillen, E., Hillenbrand, L.A., Daivd, T.J., Aigrain, S., Rebull, L., Stauffer, J., Cody, A.M., \& Queloz, D., 2017, ApJ, in press.
(https://arxiv.org/abs/1706.03084)

Girardi, L., Groenewegen, M.~A.~T., Hatziminaoglou, E., \& da Costa, L.\ 2005, \aap, 436, 895 

Girardi, L., Barbieri, M., Groenewegen, M.~A.~T., et al.\ 2012, Astrophysics and Space Science Proceedings, 26, 165 

Goodwin, S.P., Kroupa, P., Goodman, A., Burkert, A., 2007,
Protostars and Planets V, eds. B Reipurth , D. Jewitt \& K. Keil (University of Arizona Press: Tuscon), p 133-147.

Guarcello, M. G., Flaccomio, E., Micela, G., Argiroffi, C., Sciortino, S., Venuti, L., Stauffer, J., Rebull, L., Cody, A. M., 2017, \aa, 602, 10

Hauschildt, P.~H., Allard, F., \& Baron, E.\ 1999, \apj, 512, 377 

Jaschek, C. 1976, A\&A 50, 185

Jensen-Clem, R., Duev, D., Riddle, R., Salama, M., Baranec, C., Law, N., \& Kulkarni, S., 2017, PASP, submitted
(https://arxiv.org/abs/1703.08867)

Jones, B., \& Stauffer, J. 1991, AJ, 102, 1080

Jones, B., \& Cudworth, K., 1983, AJ 88, 215 

Kamezaki, T., Imura, K., Omodaka, T., et al.\ 2014, \apjs, 211, 18 

Khalaj, P., \& Baumgardt, H.\ 2013, \mnras, 434, 3236

Klein-Wassink, W. 1927, Publ. Kapteyn Astron. Lab. Groningen 41

Kraus, A. \& Hillenbrand, L., 2007, AJ, 134, 6.

Kounkel, M., Hartmann, L., Tobin, J.~J., et al.\ 2016, \apj, 821, 8 

Lafreniere et al. 2007, ApJ, 660, 770

Laher, R. Aperture Photometry Tool. Computer software. Aperture Photometry Tool. Vers. 2.5.1. Spitzer Science Center, 2015. Web. 13 July 2015.

Laher, R. et al., Publications of the Astronomy Society of the Pacific, Vol. 124, No. 917 (July 2012), pp. 764-781

Lanza, A.~F., Flaccomio, E., Messina, S., et al.\ 2016, \aap, 592, A140 


Law, N.~M., Baranec, C., \& Riddle, R.~L.\ 2014, \procspie, 9148, 91480A 

Law, N.~M., Morton, T., Baranec, C., et al.\ 2014, \apj, 791, 35

van Leeuwen, F.\ 1983, Ph.D.~Thesis,

van Leeuwen, F., Evans, D.~W., De Angeli, F., et al.\ 2017, \aap, 599, A32 

Lodieu, N., Deacon, N.~R., \& Hambly, N.~C.\ 2012, \mnras, 422, 1495

Mart{\'{\i}}n, E.~L., Brandner, W., Bouvier, J., et al.\ 2000, \apj, 543, 299 

Melis et al. 2014, Science 345, 1029

Mermilliod et al 1992, AA 265, 513  

Mermilliod et al 1997, AA 320, 74.

Mermilliod \& Mayor 1999, AA 352, 479

Offner, S.~S.~R., Kratter, K.~M., Matzner, C.~D., Krumholz, M.~R., \& Klein, R.~I.\ 2010, \apj, 725, 1485

Patience, J., Ghez, A.~M., Reid, I.~N., \& Matthews, K.\ 2002, \aj, 123, 1570

Peterson, D.~M., \& White, N.~M.\ 1984, \aj, 89, 824 

Peterson, D.~M., Baron, R., Dunham, E.~W., et al.\ 1989, \aj, 98, 2156

Pinfield, D.~J., Dobbie, P.~D., Jameson, R.~F., et al.\ 2003, \mnras, 342, 1241 

Raboud \& Mermilliod 1998, AA 329, 101

Raghavan, D., McAlister, H.~A., Henry, T.~J., et al.\ 2010, \apjs, 190, 1 

Rebull, L.~M., Stauffer, J.~R., Bouvier, J., Cody, A. M., Hillenbrand, L. A., Soderblom, D. R., Valenti, J., Barrado, D., Bouy, H., Ciardi, D., et al. \ 2016, \aj, 152, 113 

Rebull, L.~M., Stauffer, J.~R., L.A. Hillenbrand, A.M. Cody, J. Bouvier,
D.R. Soderblom, M. Pinsonneault, 2017, \apj, 839, 92 

Reipurth, B., Clarke, C.~J., Boss, A.~P., et al.\ 2014, Protostars and Planets VI, 267

Richichi, A., Chen, W.~P., Cusano, F., et al.\ 2012, \aap, 541, A96 

Riddle, R.~L., Hogstrom, K., Papadopoulos, A., Baranec, C., \& Law, N.~M.\ 2014, \procspie, 9152, 91521E

Riddle, R.~L., Tokovinin, A., Mason, B.~D., et al.\ 2015, \apj, 799, 4

Salama, M., Baranec, C., Jensen-Clem, R., et al.\ 2016, \procspie, 9909, 99091A 
(https://arxiv.org/pdf/1608.02985.pdf) 

Sarro, L.~M., Bouy, H., Berihuete, A., et al.\ 2014, \aap, 563, A45 

Shu, F.~H., Adams, F.~C., \& Lizano, S.\ 1987, \araa, 25, 23

Siess, L., Dufour, E., \& Forestini, M.\ 2000, \aap, 358, 593

Sousa, A.~P., Alencar, S.~H.~P., Bouvier, J., et al.\ 2016, \aap, 586, A47 

Stauffer, J.~R., Hartmann, L., Soderblom, D.~R., \& Burnham, N.\ 1984, \apj, 280, 202

Stauffer, J.~R., Hartmann, L.~W., Fazio, G.~G., et al.\ 2007, \apjs, 172, 663


Venuti, L., Bouvier, J., Cody, A.~M., et al.\ 2017, \aap, 599, A23

Wang, P.~F., Chen, W.~P., Lin, C.~C., et al.\ 2014, \apj, 784, 57 

Ziegler, C., Law, N.~M., Morton, T., et al.\ 2017, \aj, 153, 66

\appendix 

%

\begin{deluxetable}{c c r r r r r r c r r r r r}   
\tablewidth{0pt}
\tabletypesize{\tiny}       
\tablecaption{Characteristics of Observed Targets
\label{tab:sources}}   
\tablehead{  \colhead{Cluster} &\colhead{Target\tablenotemark{1}} & \colhead{R.A.} & \colhead{Dec.} & \colhead{V} & \colhead{I} & \colhead{J} & \colhead{K} & \colhead{Robo-AO multiplicity} & \colhead{FWHM\tablenotemark{2}} &  \multicolumn{4}{c}{Robo-AO Contrast Distribution}\\
& &  &  &  &  &  &  &&& \colhead{0.5"} & \colhead{1"} & \colhead{2"} & \colhead{3"} \\
& &\multicolumn{2}{c}{[degrees]} & \multicolumn{4}{c}{[mag]} & & \colhead{[pix]} & \multicolumn{4}{c}{[$\Delta$mag]}
}
\startdata    
Pleiades&AKIV 314	&	58.53719	&	24.33364	&	12.48	&	11.23	&	10.38	&	9.72	&	detected binary	&	9.1	&	2.92	&	4.47	&	5.96	&	6.22	\\
        &BPL 167	&	57.03671	&	25.02939	&	16.19	&	13.65	&	12.20	&	11.25	&	single	&	6.5	&	2.05	&	3.03	&	3.58	&	3.62	\\
        &BPL 273	&	58.21706	&	25.17400	&	16.27	&	13.73	&	12.38	&	11.53	&	likely single	&	4.8	&	1.68	&	2.85	&	3.32	&	3.34	\\
        &DH 027	&	53.29373	&	22.52204	&	15.55	&	13.36	&	12.12	&	11.25	&	single	&	6.0	&	2.06	&	3.27	&	3.81	&	3.86	\\
        &DH 056	&	54.10133	&	22.62379	&	14.14	&	12.37	&	11.35	&	10.56	&	detected binary	&	6.5	&	1.84	&	2.99	&	3.77	&	3.90	\\
        &DH 065	&	54.21886	&	23.26258	&	13.66	&	12.25	&	11.38	&	10.63	&	single	&	7.3	&	2.18	&	3.33	&	4.23	&	4.36	\\
        &DH 146	&	55.14717	&	20.96581	&	15.99	&	13.59	&	12.19	&	11.30	&	single	&	16.9	&	1.67	&	2.85	&	3.61	&	3.69	\\
        &DH 156	&	55.21359	&	23.59843	&	12.79	&	11.57	&	10.74	&	10.08	&	single	&	6.1	&	1.81	&	2.99	&	4.32	&	4.64	\\
\enddata
\tablenotetext{1}{In the Pleiades, the naming convention follows that 
in Stauffer et al (2007) for the stars from that list.  For other stars   
from Bouy et al (2015), we generally chose to use the name from 
the first published survey that included the star as a likely Pleiades member.
Names are SIMBAD-compliant if preceded by "Cl$*$ Melotte 22"
with the exception of those noted as "s" which come from the catalog of
Sarro et al. (2014) - \url{http://vizier.cfa.harvard.edu/viz-bin/VizieR?-source=J/A+A/563/A45}.
In Praesepe, we attempted to follow the standard convention of
referring to the star by the name given to it in the first paper which
ascribed cluster membership.  
Names are SIMBAD-compliant if preceded by "Cl$*$ NGC 2632"
with the exception of those noted as "AD", a nomenclature originating in
Kraus \& Hillenbrand (2007) to refer to stars first identified as Praesepe
members by Adams et al. (2002).
In NGC 2264, the naming convention is that of the Stauffer et al.\ Spitzer 
variability survey (\url{https://irsa.ipac.caltech.edu/data/SPITZER/CSI2264/}).  
Names are SIMBAD-compliant as recorded.
}
\tablenotetext{2}{The Robo-AO platescale is sampled at 21.55 milli-arcsec/pixel}
\end{deluxetable}

			
%

\begin{deluxetable}{c c r r r r r}   
\tablewidth{0pt}
\tabletypesize{\tiny}    
\tablenum{2}   
\tablecaption{Basic Characteristics of Detected Binaries
\label{tab:bin}}   
\tablehead{  \colhead{Cluster} & \colhead{Binary} & \colhead{Significance} & \colhead{Projected Separation} & \colhead{Position Angle} & \colhead{Optical Brightness Ratio} & \colhead{Mass Ratio} \\
 & & $\sigma$& [arcseconds]& [degrees]& [$\Delta$mag] & [$q=m_2/m_1$]
}
\startdata    
Pleiades&	AK IV-314&9.7&	1.03	&	121.2	&	2.37$\pm$0.05	& 0.53\\
	&	DH 056	&10.6&	2.52	&	244.1	&	2.78$\pm$0.10	& 0.33\\
	&	DH 193	&9.7&	0.90	&	83.9		&	1.86$\pm$0.08	& 0.54\\
	&	DH 446	&9.4&	2.28	&	64.2		&	3.39$\pm$0.17	& $<$0.23\\
	&	DH 800	&19.3 &	2.25	&	95.6	&	2.41$\pm$0.05	& 0.52\\
	&	DH 896	&11.7&	0.96	&	5.8	&	1.38$\pm$0.05	& 0.57\\
	&	HCG 86	&11.1&	1.21	&	185.2	&	0.75$\pm$0.04	& 0.71\\
	&	HCG 123	&8.4&	0.78	&	253.6	&	1.67$\pm$0.06	& 0.50\\
\enddata
\end{deluxetable}


\end{document}